\documentclass{article}
\usepackage{iclr2027_conference,times}

\usepackage{amsmath,amsfonts,bm}

\def\eqref#1{equation~\ref{#1}}

\def\1{\bm{1}}

\DeclareMathAlphabet{\mathsfit}{\encodingdefault}{\sfdefault}{m}{sl}
\SetMathAlphabet{\mathsfit}{bold}{\encodingdefault}{\sfdefault}{bx}{n}

\newcommand{\KL}{D_{\mathrm{KL}}}

\renewcommand{\eqref}[1]{(\ref{#1})}

\usepackage{amsmath,amssymb,mathtools,amsthm,bm}
\usepackage{graphicx}
\usepackage{booktabs}
\usepackage{multirow}
\usepackage{array}
\usepackage{tabularx}
\usepackage[table]{xcolor}
\usepackage{microtype}
\usepackage{enumitem}
\usepackage{placeins}
\usepackage{float}
\usepackage{wrapfig}
\usepackage{needspace}
\usepackage{caption}
\DeclareCaptionJustification{rarsfulljustified}{\frenchspacing\leftskip=0pt\rightskip=0pt\parfillskip=0pt}
\usepackage{hyperref}
\usepackage{url}

\definecolor{logicblue}{HTML}{3F6FA6}
\definecolor{logicorange}{HTML}{D97706}
\definecolor{logicgreen}{HTML}{3F8F5B}
\definecolor{logicred}{HTML}{B64E4E}
\definecolor{logiclight}{HTML}{EDF4FB}
\definecolor{adalignrow}{HTML}{EFE6D7}
\definecolor{logicgray}{HTML}{F2F3F5}
\definecolor{pendingfg}{HTML}{777777}
\hypersetup{colorlinks=true,citecolor=brown,linkcolor=red,anchorcolor=red,urlcolor=logicblue}

\newcommand{\CE}{\operatorname{CE}}
\newcommand{\Children}{\operatorname{Ch}}

\newcommand{\venueyear}[1]{\textcolor{logicorange}{\scriptsize #1}}
\newcommand{\best}[1]{\textbf{#1}}
\newcommand{\second}[1]{\underline{#1}}

\newsavebox{\maintablenumberbox}
\newcommand{\maintablenumber}[1]{%
  \begingroup
  \sbox{\maintablenumberbox}{#1}%
  \raisebox{0pt}[\ht\maintablenumberbox][\dp\maintablenumberbox]{%
    \makebox[\wd\maintablenumberbox][r]{\scalebox{1.15}{#1}}}%
  \endgroup
}

\title{Learning Multiresolution Relevance for   \\  Hierarchical Generative Retrieval}

\author{%
\textbf{Weihao Shen$^{1,2}$, Wei Chen$^{1,2}$, Fuwei Zhang$^{1}$, Guojun Liu$^{2}$}\\
\textbf{Qingsong Hua$^{2}$, Wei Lin$^{2}$, Fuzhen Zhuang$^{1}$\thanks{Corresponding author.}}\\[0.5em]
{\normalfont $^{1}$Institute of Artificial Intelligence, Beihang University, Beijing, China}\\
{\normalfont $^{2}$Meituan, Beijing, China}\\[0.3em]
{\normalfont\texttt{\{shenweihao,chenwei23,zhuangfuzhen\}@buaa.edu.cn}}
}

\iclrfinalcopy
\begin{document}
\addtocontents{toc}{\protect\setcounter{tocdepth}{-1}}
\setcounter{footnote}{1}
\maketitle
\fancyhead[L]{}

\begin{abstract}
Generative retrieval with semantic identifiers (SIDs) makes successive decisions over a document hierarchy. Relevant documents for the same query may share coarse prefixes and diverge at finer depths, with branching patterns varying across queries. These paths reveal how relevance is distributed across successive refinements, yet standard full-SID supervision treats them as separate training targets. To make this allocation explicit, we formulate multiresolution relevance as consistent conditional distributions induced by a single document-level relevance measure across the SID hierarchy. We introduce \textbf{RARS}, \textbf{R}esolution-\textbf{A}ligned \textbf{R}elevance \textbf{S}upervision, which uses the resulting refinement-level distributions to supervise a shared query representation. RARS aggregates document relevance over prefixes and trains a prefix-conditioned predictor to allocate relevance among sibling branches. All relevance-bearing children participate in local competition, and each local loss is weighted by the relevance mass reaching its parent. This objective trains the query encoder to capture both the coarse structure shared by relevant documents and their finer branch allocations. The predictor is discarded after training, preserving standard autoregressive retrieval at inference. Experiments on three multilingual ESCI locales show consistent improvements over matched full-SID training under autoregressive decoding. RARS also outperforms grouped soft-target, decoder soft-target, and sampled-tree supervision under a common retrieval rule. The gains persist across alternative identifier structures and relevance definitions.
Code is available at: \href{https://github.com/Nevaeh7/RARS}{\textcolor{blue}{\texttt{https://github.com/Nevaeh7/RARS}}}.
\end{abstract}

\section{Introduction}

\label{sec:introduction}

\suppressfloats[t]

\begin{figure}[t]
    \centering
    \includegraphics[width=\linewidth]{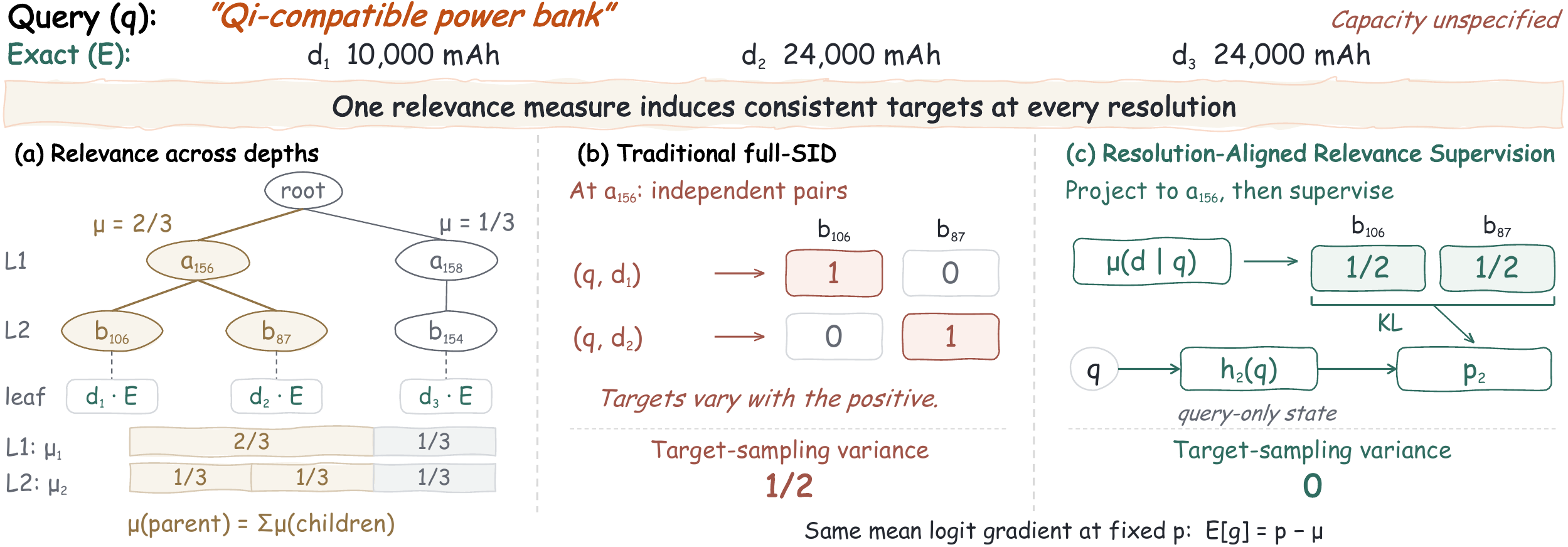}
    \caption{\textbf{Relevance across SID resolutions.}
    (a) Aggregating document relevance over prefixes preserves total mass.
    (b) Full-SID supervision presents each positive as a separate path.
    (c) RARS learns the conditional allocation $[1/2,1/2]$ at the shared prefix
    from a shared query representation.}
    \label{fig:logictrie-intro}
\end{figure}

Generative retrieval casts retrieval as sequence generation over document
identifiers
\citep{tay2022dsi,wang2022nci,bevilacqua2022seal}.
Semantic identifiers (SIDs) further organize documents into structured codes
whose prefixes capture progressively finer distinctions
\citep{wang2022nci,sun2023tokenize,rajput2023tiger,zhang2025hiergr,liu2026catid}.
A shallow prefix identifies a coarse region of the corpus, and subsequent
tokens progressively refine that region toward an individual document.
SID-based retrieval is therefore inherently multiresolution: each generation
step resolves the query over a finer corpus partition, and an incorrect prefix
can exclude an entire relevant subtree from constrained search
\citep{wang2022nci,zeng2024ripor}.
Because the complete SID also serves as the retrieval address, full-SID
supervision is a natural training target.

This supervision is limited for queries with multiple relevant documents.
Relevant documents may share a prefix and diverge only at a later SID depth,
so a single query can support several competing child branches under the same
parent. Figure~\ref{fig:logictrie-intro} illustrates this case with the query
``Qi-compatible power bank.'' The query specifies charging compatibility while
leaving capacity unspecified, allowing several Exact products to share coarse
SID structure and separate at finer retrieval resolutions. At one depth,
relevance may be concentrated on a single branch; at a subsequent depth, the
same relevance mass may divide among several children. Standard full-SID
training presents these products as separate root-to-leaf targets
\citep{tang2024gr2,li2024ltrgr,tang2024listwise}, so the same query is
associated with different child decisions at a shared parent. At the query level, these paths induce a relevance distribution over competing
refinements, but this distribution is not represented explicitly during
training. Full-path supervision therefore leaves the local relevance structure
of intermediate SID decisions implicit. We refer to this granularity gap as a
supervision--resolution mismatch.

Recent work has incorporated ranking objectives, prefix-level preferences,
graded relevance, relevance-aware identifiers, and multi-positive supervision
\citep{zhou2023rlgr,zhang2025merge,zeng2024ripor,zhang2026calir}.
These directions establish the value of relevance-aware learning in generative
retrieval, but leave open how document-level relevance should be represented
when retrieval proceeds through nested corpus partitions and local branch
decisions. We address this question with a multiresolution view of relevance.
For each query, document-level relevance defines a distribution over relevant
leaves of the SID hierarchy. Projecting this distribution onto each depth
yields a relevance measure over the corresponding corpus partition. Because
the partitions are nested, parent mass equals the total mass of its children,
and conditioning child mass on the parent gives a local relevance distribution
for each refinement. The targets at different depths are therefore consistent
views of the same query relevance at progressively finer hierarchical retrieval
resolutions.

To learn these multiresolution relevance targets, we introduce
RARS, Resolution-Aligned Relevance Supervision. RARS uses a shared query
representation to predict the relevance allocation at each supported refinement
through depth-specific projections and prefix-conditioned child scores. Each
local target retains all relevance-bearing branches, including those that diverge
only at finer SID depths, thereby preserving the query's multiresolution
relevance structure. The loss weights each refinement by the relevance mass
reaching its parent and includes competing legal siblings in local
normalization. This design couples supervision across the query's relevant paths
through a single encoder representation, requiring the representation to encode
branch allocations across successive resolutions. The multiresolution objective is optimized alongside the full-SID objective.
The hierarchical predictor is discarded after training, preserving standard
autoregressive retrieval.

We evaluate RARS on the English, Spanish, and Japanese ESCI product-search
benchmarks \citep{reddy2022esci}. Controlled experiments separate the effects
of distributional supervision, hierarchical prediction, and retrieval
scoring. RARS improves over matched full-SID training under standard
autoregressive decoding across all three locales and remains stronger than
grouped soft-target, decoder soft-target, and sampled-tree controls. The gains
persist across alternative SID constructions and relevance definitions.
Further analyses show that relevance branching is common at coarse and
intermediate SID depths, while incomplete judgments can obscure this
structure.

Our contributions are threefold:

\textbf{(i)} We formulate multiresolution relevance for generative retrieval.
A single document-level relevance measure induces consistent conditional
targets across SID refinements, and a refinement-entropy profile identifies
where observed relevance supports competing branches.

\textbf{(ii)} We propose RARS to learn branch allocations through a
shared query representation. Prefix-conditioned prediction and parent-mass
weighting supervise relevance-bearing refinements, while local sibling
competition encourages the encoder to distinguish competing child branches.
The predictor is used only during training, without modifying autoregressive
retrieval at inference.

\textbf{(iii)} Controlled experiments on three multilingual ESCI locales
establish gains over grouped soft-target, decoder soft-target, and sampled-tree
supervision. Further comparisons verify improvements under
autoregressive retrieval and across alternative SID constructions and
relevance definitions.

\section{Related Work}
\label{sec:related}

\textbf{Generative retrieval and structured identifiers.}
Generative retrieval maps queries directly to document or entity identifiers
~\citep{decao2021genre,tay2022dsi}. NCI employs hierarchical identifiers
with prefix-aware decoding, while SEAL generates document
substrings~\citep{wang2022nci,bevilacqua2022seal}. Learned tokenization,
semantic trees, and residual quantization provide alternative constructions of
structured identifiers with progressively finer semantic organization
~\citep{sun2023tokenize,si2023semantic,rajput2023tiger,jin2024lmindexer}.
For product search, Hi-Gen, MERGE, and CAT-ID$^2$ incorporate category
structure or query relevance into identifier construction
~\citep{wu2024higen,zhang2025merge,liu2026catid}. CaLIR further introduces
category-guided latent reasoning and query-conditioned
tries~\citep{zhang2026calir}. We build on CaLIR and focus on a complementary
question: how document-level relevance should directly supervise local decisions
across successive refinements of the SID hierarchy at multiple retrieval
resolutions.

\textbf{Relevance-aware supervision.}
Listwise ranking methods model relative relevance across documents:
ListNet matches score-induced distributions, while ListMLE models ranking
likelihood~\citep{cao2007listnet,xia2008listmle}. In generative retrieval,
LTRGR learns from document rankings, RIPOR from pairwise preferences over
prefixes, and GR$^2$ from graded relevance~\citep{li2024ltrgr,zeng2024ripor,tang2024gr2}.
Relevance feedback and rank distillation align identifier generation
with retrieval quality~\citep{zhou2023rlgr,li2024dgr}.
Multi-positive objectives supervise representations using multiple relevant
items~\citep{khosla2020supcon}.
In particular, CaLIR averages over positive category prototypes at each
level~\citep{zhang2026calir}. RARS aggregates document relevance over SID
prefixes to characterize how a query's relevance is distributed at each successive refinement.
The projection preserves relevance mass across depths, while the conditional
targets retain the relative relevance among child branches. RARS learns these allocations
through a shared query representation across relevant paths, using
prefix-conditioned prediction and parent-mass weighting.

\textbf{Hierarchical prediction.}
Label trees factorize predictions along root-to-leaf
paths~\citep{bengio2010labeltrees,wydmuch2018noregret}, while hierarchical
classifiers incorporate this structure into probabilities and
losses~\citep{ramaswamy2015hierarchical,bertinetto2020making}.
Selective hierarchical classification can return internal nodes when
predictions are uncertain~\citep{goren2024hierarchical}. RARS uses the SID
hierarchy to define query-conditioned relevance targets for refinement
decisions. Its predictor supervises the query encoder across relevant
branches during training, while retrieval retains the full-SID decoder.
The refinement-entropy profile characterizes how observed relevance is distributed
across the hierarchy. Confidence calibration concerns the reliability of
predictions~\citep{guo2017calibration}.
\section{Preliminaries}
\label{sec:preliminaries}

\textbf{Semantic identifiers.}
Let $\mathcal D$ denote a document corpus, and let
$s(d)=(s_1(d),\ldots,s_L(d))$ be the length-$L$ semantic identifier (SID)
of document $d\in\mathcal D$
\citep{wang2022nci,sun2023tokenize,rajput2023tiger}.
We denote its depth-$\ell$ prefix by
$\pi_\ell(d)=(s_1(d),\ldots,s_\ell(d))$, with
$\pi_0(d)=\varnothing$.
Let $\mathcal P_\ell=\{\pi_\ell(d):d\in\mathcal D\}$ be the set of valid
prefixes at depth $\ell$, with $\mathcal P_0=\{\varnothing\}$, and let
$\mathrm{Ch}(u)$ denote the valid children of
$u\in\mathcal P_{\ell-1}$.
The resulting SID trie induces a sequence of nested corpus partitions:
shallow prefixes represent coarse document groups, while deeper prefixes
progressively refine them toward individual documents.

\textbf{Autoregressive generative retrieval.}
Given a query $q$, a generative retriever models a document SID as a sequence
of refinement decisions and factorizes its probability along the SID path as
\begin{equation}
    p_\theta(s(d)\mid q)
    =
    \prod_{\ell=1}^{L}
    p_\theta\!\left(
        s_\ell(d)\mid q,s_{<\ell}(d)
    \right),
    \label{eq:ar}
\end{equation}
where $s_{<\ell}(d)=(s_1(d),\ldots,s_{\ell-1}(d))$ denotes the previously
generated prefix.
Each factor in Eq.~\eqref{eq:ar} corresponds to a local refinement in the
SID trie: conditioned on the query and current prefix, the model selects
among valid child branches.
Under trie-constrained decoding, invalid continuations are excluded, so each
prefix decision determines which subset of documents remains reachable.

\textbf{Full-SID supervision.}
Let $\mathcal R$ denote the set of relevant query--document pairs.
Standard full-SID supervision treats each pair as a target path
and minimizes the maximum-likelihood objective
\begin{equation}
    \mathcal L_{\mathrm{leaf}}
    =
    -\mathbb E_{(q,d)\sim\mathcal R}
    \log p_\theta(s(d)\mid q).
    \label{eq:leaf}
\end{equation}
Each relevant pair contributes a deterministic root-to-leaf path.
For a query with multiple relevant documents, these paths may share a prefix
and therefore a common coarse decision, then diverge at a later refinement,
thereby specifying different child branches below the same parent.
The next section develops a query-conditioned relevance distribution over
these refinements.
\section{RARS: Relevance Across Retrieval Resolutions}
\label{sec:method}

\suppressfloats[t]

\begin{figure}[t]
    \centering
    \includegraphics[width=\linewidth]{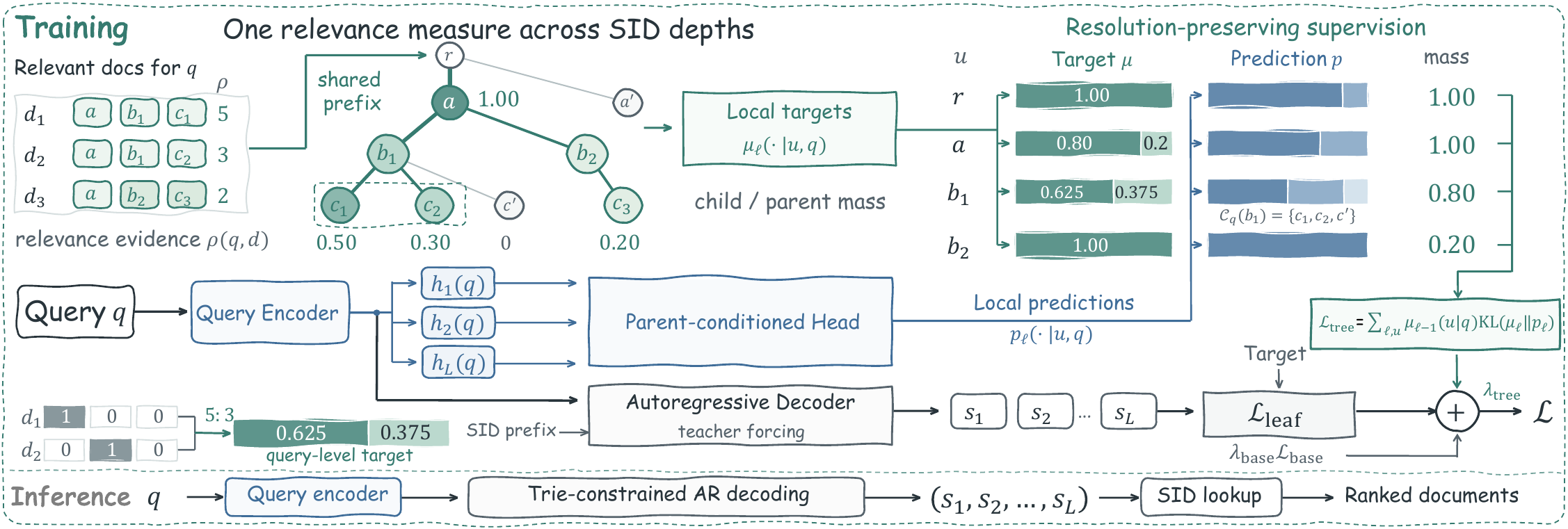}
\caption{\textbf{RARS overview.}
Document relevance is projected onto SID partitions to form conditional targets at each refinement. A query-level predictor learns these distributions with parent-mass-weighted supervision, while the full-SID decoder is retained and used alone at inference.}
    \label{fig:rars-method}
\end{figure}

RARS trains a query representation to predict how relevance is allocated
across successive SID refinements (Figure~\ref{fig:rars-method}).
Document relevance defines the target mass on the leaves, and the SID
partitions determine its allocation to prefixes and their children.
A prefix-conditioned predictor learns these local allocations from the
shared query encoder. Parent-mass weighting determines the contribution of
each refinement to training, and local sibling competition supplies the
candidate alternatives. The full-SID decoder retains its original objective
and retrieval procedure.

\subsection{Relevance Across Identifier Resolutions}
\label{sec:relevance-resolution}

For a query $q$, let $\mathcal D_q^+$ denote the set of observed relevant
documents, and let $\rho(q,d)\geq 0$ denote the relevance evidence associated
with $d\in\mathcal D_q^+$. We define the normalized document-level relevance
measure as
$\mu(d\mid q)=\rho(q,d)/\sum_{d'\in\mathcal D_q^+}\rho(q,d')$.
Binary relevance assigns equal mass to all observed positives, while graded
judgments or continuous relevance scores induce non-uniform masses.

The SID hierarchy induces a corpus partition at each depth. At depth $\ell$,
we obtain the relevance mass of a prefix by aggregating the masses of its
relevant descendant documents:
\begin{equation}
    \mu_\ell(z\mid q)
    =
    \sum_{d\in\mathcal D_q^+}
    \mu(d\mid q)\,
    \mathbf 1[\pi_\ell(d)=z],
    \qquad z\in\mathcal P_\ell .
    \label{eq:projection}
\end{equation}
Hence, $\mu_\ell$ gives the relevance distribution induced by the same
document-level measure at SID resolution $\ell$.
Since SID partitions are nested, the projected distributions are consistent
across depths. For each parent prefix $u\in\mathcal P_{\ell-1}$, its relevance
mass equals the total mass of its children. When
$\mu_{\ell-1}(u\mid q)>0$, the normalized child masses define the conditional
distribution
$\mu_\ell(v\mid u,q)=\mu_\ell(v\mid q)/\mu_{\ell-1}(u\mid q)$.
This distribution specifies the allocation of relevance among the child
branches of $u$. Consequently, the supervision at different depths is induced
by a common document-level relevance measure and remains consistent with the
hierarchical structure.

We quantify how relevance is distributed across competing refinements.
Let $D\sim\mu(\cdot\mid q)$ and $Z_\ell=\pi_\ell(D)$ denote the SID prefix
of $D$ at depth $\ell$. The refinement ambiguity at depth $\ell$ is defined as
\begin{equation}
    \Delta_\ell(q)
    =
    H_\mu(Z_\ell\mid Z_{\ell-1},q)
    =
    \sum_{u\in\mathcal P_{\ell-1}}
    \mu_{\ell-1}(u\mid q)\,
    H\!\left(\mu_\ell(\cdot\mid u,q)\right).
    \label{eq:ambiguity}
\end{equation}
The quantity $\Delta_\ell(q)$ is zero when every relevance-bearing parent
assigns all of its mass to a single child, and increases as relevance is
distributed over multiple sibling branches. It is determined solely by the
observed relevance evidence and the SID hierarchy, and does not depend on model
predictions. The profile $\{\Delta_\ell(q)\}_{\ell=1}^{L}$ therefore
characterizes the SID depths at which observed relevance remains distributed
across competing refinements.

\subsection{Query-Level Hierarchical Prediction}
\label{sec:hierarchical-prediction}

The projected targets specify what should be predicted at each refinement.
We parameterize these predictions using a representation shared across
relevant documents of the query.
Let $\bar h(q)$ be the masked mean of the query encoder's final token
representations.
A depth-specific projection produces
$h_\ell(q)=W_\ell\bar h(q)\in\mathbb R^r$.
For a parent prefix $u=(u_1,\ldots,u_{\ell-1})$, we encode its path as
$c(u)=\sum_{j<\ell}b_{j,u_j}$ using position-specific prefix embeddings,
with $c(\varnothing)=0$.
The score of child $v$ is
\begin{equation}
    a_\ell(v\mid u,q)
    =
    \frac{
        e_{\ell,v}^{\top}
        \left[
            h_\ell(q)\odot(1+\tanh c(u))
        \right]
    }{\sqrt r}
    +\beta_{\ell,v}.
    \label{eq:tree-score}
\end{equation}
Conditioning on $c(u)$ allows the same token value to receive different
scores under different prefixes, while $h_\ell(q)$ preserves a common
query-level state across the relevant paths.

The local predictive distribution is obtained by normalizing over a candidate
set $\mathcal C_q(u)$:
\begin{equation}
    p_\ell(v\mid u,q)
    =
    \frac{
        \exp\!\left(a_\ell(v\mid u,q)/\tau_p\right)
    }{
        \sum_{c\in\mathcal C_q(u)}
        \exp\!\left(a_\ell(c\mid u,q)/\tau_p\right)
    } .
    \label{eq:tree-posterior}
\end{equation}
The candidate set at each relevance-bearing parent includes every
relevance-bearing child and competing legal siblings.
This local normalization maintains competition among children of the same
parent without requiring a softmax over the full identifier space.
We use $r=64$ and $\tau_p=1$, with up to 32 high-scoring legal sibling
negatives included alongside all relevance-bearing children.

\subsection{Resolution-Preserving Objective}
\label{sec:resolution-objective}

We train the hierarchical predictor to match the projected relevance
distribution at every relevance-bearing refinement, aligning predictions
with the relevance structure induced by the SID hierarchy:
\begin{equation}
    \mathcal L_{\mathrm{tree}}(q)
    =
    \sum_{\ell=1}^{L}
    \sum_{u\in\mathcal P_{\ell-1}}
    \mu_{\ell-1}(u\mid q)\,
    D_{\mathrm{KL}}\!\left(
        \mu_\ell(\cdot\mid u,q)
        \,\middle\|\,
        p_\ell(\cdot\mid u,q)
    \right).
    \label{eq:tree-loss}
\end{equation}
The parent mass $\mu_{\ell-1}(u\mid q)$ measures how much query relevance
reaches the corresponding refinement.
It weights each local loss by the relevance carried through that parent,
preserving the contribution of each branch to the query-level objective.
All relevance-bearing children remain in $\mathcal C_q(u)$, so each local
loss supervises the allocation among supported branches in the presence of
selected legal sibling negatives. The shared query representation receives
these constraints from every relevance-bearing parent.
Appendix~\ref{app:path-consistency} derives the full-path consistency of the
weighting under full-child normalization and its relation to grouped
supervision.
The complete training objective is
\begin{equation}
    \mathcal L
    =
    \mathcal L_{\mathrm{leaf}}
    +
    \lambda_{\mathrm{tree}}\mathcal L_{\mathrm{tree}}
    +
    \lambda_{\mathrm{base}}\mathcal L_{\mathrm{base}}.
    \label{eq:objective}
\end{equation}
Our reference retriever is CaLIR
\citep{zhang2026calir}, which combines full-SID likelihood with category
classification and multi-positive category alignment.
We retain its original auxiliary objective
$\mathcal L_{\mathrm{base}}
=\mathcal L_{\mathrm{cls}}+\mathcal L_{\mathrm{con}}$
with $\lambda_{\mathrm{base}}=0.1$, and set
$\lambda_{\mathrm{tree}}=0.5$.
The matched base optimizes
$\mathcal L_{\mathrm{leaf}}
+0.1(\mathcal L_{\mathrm{cls}}+\mathcal L_{\mathrm{con}})$.
RARS adds $\mathcal L_{\mathrm{tree}}$ to supervise the query encoder at
the local refinement decisions.
Gradients from $\mathcal L_{\mathrm{tree}}$ update the hierarchical predictor
and the shared query encoder.
The autoregressive decoder and the remaining retriever parameters continue to
be optimized by their original objectives.
Since the hierarchical predictor is absent at inference, its effect on
retrieval is mediated entirely through the representations learned during
training.

\subsection{Supervision at Branching Prefixes}
\label{sec:relation-fullsid}

For a query with a single relevant document $d^\star$ and
$\mu(d^\star\mid q)=1$,
$\mu_\ell(z\mid q)=\mathbf 1[z=\pi_\ell(d^\star)]$ for every depth $\ell$.
All local targets therefore collapse to the same deterministic root-to-leaf
path.
For queries with multiple relevant documents, relevance can branch below a
shared prefix.
Consider one such parent, with target distribution $\bm\mu$ over its children
and model prediction $\bm p$.
Sampling $Y\sim\bm\mu$ produces the one-hot logit gradient
$\bm g_Y=\bm p-\bm e_Y$, whereas direct distributional supervision produces
$\bm g_\mu=\bm p-\bm\mu$.
For the same temperature-scaled local logits,
\begin{equation}
    \mathbb E_Y[\bm g_Y]
    =
    \bm g_\mu,
    \qquad
    \mathbb E_Y
    \left[
        \|\bm g_Y-\bm g_\mu\|_2^2
    \right]
    =
    1-\|\bm\mu\|_2^2 .
    \label{eq:gradient-variance}
\end{equation}
Relevance-weighted target sampling and explicit distributional
supervision have the same expected local gradient.
The latter removes variance from drawing a target,
while leaving other sources of optimization stochasticity unchanged.
The variance term in Eq.~\eqref{eq:gradient-variance} is the Gini impurity,
whereas Eq.~\eqref{eq:ambiguity} uses Shannon entropy.
With natural logarithms, their parent-weighted quantities satisfy
\begin{equation}
    \mathcal V_\ell(q)
    :=
    \sum_u
    \mu_{\ell-1}(u\mid q)
    \left[
        1-\|\mu_\ell(\cdot\mid u,q)\|_2^2
    \right]
    \leq
    \Delta_\ell(q).
    \label{eq:ambiguity-bound}
\end{equation}
Both quantities vanish for deterministic refinements, but they capture
different properties: $\Delta_\ell(q)$ characterizes the dispersion of
relevance across siblings, whereas $\mathcal V_\ell(q)$ measures the variance
introduced by sampling a one-hot target.
At a branching prefix, the explicit target supplies a gradient for the
entire observed relevance allocation in each update. Through
Eq.~\eqref{eq:tree-score}, these local gradients update the same query
representation across parents and depths. RARS thus couples the learning of
shared coarse prefixes with the learning of their finer relevant branches.

\subsection{Inference and Retrieval Scoring}
\label{sec:inference}

The hierarchical predictor is not used during retrieval.
Let $\mathcal S_C(q)$ denote the set of identifiers admitted by the top-$C$
category predictions of the retriever.
Category probabilities determine this candidate set but do not enter the
identifier score.
Writing $s_{L+1}=\mathrm{EOS}$, the autoregressive score is
\begin{equation}
    S_{\mathrm{AR}}(s\mid q)
    =
    \sum_{t=1}^{L+1}
    \log p_\theta(s_t\mid q,s_{<t}),
    \qquad s\in\mathcal S_C(q).
    \label{eq:inference-ar}
\end{equation}
Trie constraints assign score $-\infty$ to illegal continuations.
Equation~\eqref{eq:inference-ar} is the retrieval rule used for the matched
AR-only comparison.
We evaluate a compatibility-augmented scoring rule.
Let $f(q,d)$ measure query--catalog compatibility using catalog attributes
and lexical evidence, without relevance judgments.
For a prefix $u$ at depth $t$, define its subtree compatibility as
$F_t(u,q)=\max_{d:\pi_t(d)=u}f(q,d)$.
With a depth mask $m\in\{0,1\}^{L}$, the beam score is
\begin{equation}
    S_{\mathrm{LT}}(s\mid q)
    =
    S_{\mathrm{AR}}(s\mid q)
    +
    \lambda_{\mathrm{comp}}
    \sum_{t=1}^{L}
    m_t F_t(s_{\leq t},q),
    \qquad s\in\mathcal S_C(q).
    \label{eq:rars-score}
\end{equation}
Compatibility is added only to legal SID continuations; EOS receives no
compatibility term.
Matched comparisons under these rules distinguish training gains from
compatibility-based decoding.

\section{Experiments}
\label{sec:experiments}

We study seven research questions:
\textbf{RQ1}:~How does RARS compare with baselines,
and how do training and retrieval scoring contribute?
\textbf{RQ2}:~How does RARS compare with matched training controls
under a common retrieval rule?
\textbf{RQ3}:~How is relevance distributed across SID resolutions,
and how does incomplete evidence affect its fidelity?
\textbf{RQ4}:~Are the gains robust to identifier geometry,
training evidence, and objective design?
\textbf{RQ5}:~Does multiresolution scoring improve local prediction
alignment and prefix coverage in final rankings?
\textbf{RQ6}:~Which retrieval resolutions improve effectiveness?
\textbf{RQ7}:~How can we intuitively understand the advantage of RARS?

\begin{table}[t]
\centering
\resizebox{\textwidth}{!}{%
\begin{tabular}{lrrrrrrrrrrrrrrr}
\toprule
& \multicolumn{5}{c}{ESCI-US} & \multicolumn{5}{c}{ESCI-ES} & \multicolumn{5}{c}{ESCI-JP}\\
\cmidrule(lr){2-6}\cmidrule(lr){7-11}\cmidrule(lr){12-16}
Model & R@5 & R@10 & R@100 & N@10 & N@100 & R@5 & R@10 & R@100 & N@10 & N@100 & R@5 & R@10 & R@100 & N@10 & N@100\\
\midrule
\multicolumn{16}{l}{\emph{Sparse retrieval}}\\
BM25 \venueyear{(FnT IR'09)} & \maintablenumber{4.02}&\maintablenumber{5.84}&\maintablenumber{14.08}&\maintablenumber{5.64}&\maintablenumber{8.11} &\maintablenumber{4.13}&\maintablenumber{6.15}&\maintablenumber{16.21}&\maintablenumber{7.45}&\maintablenumber{10.20} &\maintablenumber{4.70}&\maintablenumber{7.96}&\maintablenumber{16.73}&\maintablenumber{8.63}&\maintablenumber{11.25}\\
\midrule
\multicolumn{16}{l}{\emph{Dense retrieval}}\\
DPR \venueyear{(EMNLP'20)} &\maintablenumber{5.54}&\maintablenumber{8.93}&\maintablenumber{29.30}&\maintablenumber{8.17}&\maintablenumber{15.55} &\maintablenumber{5.24}&\maintablenumber{7.08}&\maintablenumber{25.27}&\maintablenumber{8.52}&\maintablenumber{14.33} &\maintablenumber{4.18}&\maintablenumber{7.26}&\maintablenumber{23.84}&\maintablenumber{8.98}&\maintablenumber{14.73}\\
MPNet \venueyear{(NeurIPS'20)} &\maintablenumber{2.76}&\maintablenumber{4.58}&\maintablenumber{15.30}&\maintablenumber{4.12}&\maintablenumber{9.83} &\maintablenumber{2.53}&\maintablenumber{4.02}&\maintablenumber{13.27}&\maintablenumber{5.87}&\maintablenumber{8.91} &\maintablenumber{1.13}&\maintablenumber{1.97}&\maintablenumber{5.58}&\maintablenumber{1.79}&\maintablenumber{2.82}\\
Sentence-T5 \venueyear{(ACL'22)} &\maintablenumber{4.63}&\maintablenumber{6.97}&\maintablenumber{24.59}&\maintablenumber{6.84}&\maintablenumber{11.58} &--&--&--&--&-- &--&--&--&--&--\\
Sentence-mT5 \venueyear{(COLING'24)} &--&--&--&--&-- &\maintablenumber{5.44}&\maintablenumber{8.68}&\maintablenumber{30.04}&\maintablenumber{8.42}&\maintablenumber{15.09} &\maintablenumber{5.69}&\maintablenumber{8.73}&\maintablenumber{29.01}&\maintablenumber{9.19}&\maintablenumber{15.70}\\
BGE-M3 \venueyear{(ACL'24)} &\maintablenumber{6.59}&\maintablenumber{9.82}&\maintablenumber{32.29}&\maintablenumber{9.91}&\maintablenumber{16.77} &\maintablenumber{5.71}&\maintablenumber{9.07}&\maintablenumber{28.75}&\maintablenumber{9.76}&\maintablenumber{16.02} &\maintablenumber{5.14}&\maintablenumber{8.92}&\maintablenumber{27.93}&\maintablenumber{10.12}&\maintablenumber{17.02}\\
LaSER \venueyear{(SIGIR'26)} &\maintablenumber{6.52}&\maintablenumber{10.23}&\maintablenumber{35.37}&\maintablenumber{8.19}&\maintablenumber{15.74} &\maintablenumber{5.28}&\maintablenumber{8.40}&\maintablenumber{32.22}&\maintablenumber{8.69}&\maintablenumber{16.16} &\maintablenumber{5.88}&\maintablenumber{8.78}&\maintablenumber{28.66}&\maintablenumber{9.28}&\maintablenumber{15.53}\\
\midrule
\multicolumn{16}{l}{\emph{Generative retrieval}}\\
DSI$_{\mathrm{naive}}$ \venueyear{(NeurIPS'22)} &\maintablenumber{0.42}&\maintablenumber{1.74}&\maintablenumber{2.03}&\maintablenumber{0.32}&\maintablenumber{0.99} &\maintablenumber{0.28}&\maintablenumber{0.94}&\maintablenumber{1.83}&\maintablenumber{0.77}&\maintablenumber{1.60} &\maintablenumber{0.19}&\maintablenumber{0.24}&\maintablenumber{1.35}&\maintablenumber{0.21}&\maintablenumber{1.11}\\
DSI$_{\mathrm{semantic}}$ \venueyear{(NeurIPS'22)} &\maintablenumber{3.74}&\maintablenumber{6.02}&\maintablenumber{20.69}&\maintablenumber{6.24}&\maintablenumber{10.60} &\maintablenumber{4.27}&\maintablenumber{7.13}&\maintablenumber{21.63}&\maintablenumber{9.92}&\maintablenumber{14.23} &\maintablenumber{4.08}&\maintablenumber{7.40}&\maintablenumber{23.90}&\maintablenumber{9.15}&\maintablenumber{14.58}\\
TIGER \venueyear{(NeurIPS'23)} &\maintablenumber{4.79}&\maintablenumber{7.84}&\maintablenumber{25.98}&\maintablenumber{7.24}&\maintablenumber{12.68} &\maintablenumber{3.83}&\maintablenumber{7.31}&\maintablenumber{25.03}&\maintablenumber{8.72}&\maintablenumber{14.51} &\maintablenumber{4.67}&\maintablenumber{7.98}&\maintablenumber{25.89}&\maintablenumber{9.67}&\maintablenumber{15.52}\\
Hi-Gen \venueyear{(ICDM'24)} &\maintablenumber{3.13}&\maintablenumber{4.97}&\maintablenumber{15.91}&\maintablenumber{5.25}&\maintablenumber{8.55} &\maintablenumber{2.97}&\maintablenumber{5.65}&\maintablenumber{20.23}&\maintablenumber{7.06}&\maintablenumber{11.85} &\maintablenumber{3.36}&\maintablenumber{6.40}&\maintablenumber{21.66}&\maintablenumber{7.73}&\maintablenumber{12.86}\\
LTRGR \venueyear{(AAAI'24)} &\maintablenumber{2.88}&\maintablenumber{4.71}&\maintablenumber{13.70}&\maintablenumber{4.48}&\maintablenumber{7.82} &\maintablenumber{5.13}&\maintablenumber{8.23}&\maintablenumber{27.92}&\maintablenumber{9.26}&\maintablenumber{15.42} &\maintablenumber{5.24}&\maintablenumber{8.37}&\maintablenumber{26.79}&\maintablenumber{10.01}&\maintablenumber{16.93}\\
RIPOR \venueyear{(WWW'24)} &\maintablenumber{5.05}&\maintablenumber{8.35}&\maintablenumber{26.80}&\maintablenumber{7.92}&\maintablenumber{13.75} &\maintablenumber{3.49}&\maintablenumber{7.12}&\maintablenumber{23.03}&\maintablenumber{7.94}&\maintablenumber{13.82} &\maintablenumber{5.02}&\maintablenumber{8.18}&\maintablenumber{27.87}&\maintablenumber{9.92}&\maintablenumber{17.28}\\
MERGE \venueyear{(ACL'25)} &\maintablenumber{5.68}&\maintablenumber{9.26}&\maintablenumber{29.74}&\maintablenumber{9.05}&\maintablenumber{15.17} &\maintablenumber{5.80}&\maintablenumber{9.25}&\maintablenumber{30.86}&\maintablenumber{10.20}&\maintablenumber{17.45} &\maintablenumber{5.21}&\maintablenumber{8.90}&\maintablenumber{27.64}&\maintablenumber{10.95}&\maintablenumber{17.60}\\
CaLIR \venueyear{(arXiv'26)} &\maintablenumber{7.22}&\maintablenumber{11.75}&\maintablenumber{36.15}&\maintablenumber{10.75}&\maintablenumber{18.14} &\maintablenumber{6.06}&\maintablenumber{10.43}&\maintablenumber{34.83}&\maintablenumber{12.15}&\maintablenumber{20.03} &\maintablenumber{5.57}&\maintablenumber{9.64}&\maintablenumber{\second{31.71}}&\maintablenumber{11.74}&\maintablenumber{18.85}\\
CAT-ID\textsuperscript{2} \venueyear{(WSDM'26)} &\maintablenumber{6.11}&\maintablenumber{8.89}&\maintablenumber{29.03}&\maintablenumber{8.96}&\maintablenumber{14.53} &\maintablenumber{5.70}&\maintablenumber{9.34}&\maintablenumber{31.44}&\maintablenumber{10.68}&\maintablenumber{18.01} &\maintablenumber{5.36}&\maintablenumber{8.87}&\maintablenumber{28.82}&\maintablenumber{10.34}&\maintablenumber{17.14}\\
FORGE \venueyear{(KDD'26)} &\maintablenumber{5.59}&\maintablenumber{9.26}&\maintablenumber{35.11}&\maintablenumber{8.50}&\maintablenumber{16.21} &\maintablenumber{4.87}&\maintablenumber{8.29}&\maintablenumber{28.42}&\maintablenumber{9.99}&\maintablenumber{16.66} &\maintablenumber{3.88}&\maintablenumber{6.58}&\maintablenumber{24.06}&\maintablenumber{8.27}&\maintablenumber{14.22}\\
HGRec \venueyear{(ICML'26)} &\maintablenumber{6.70}&\maintablenumber{11.41}&\maintablenumber{34.61}&\maintablenumber{10.19}&\maintablenumber{17.32} &\maintablenumber{5.76}&\maintablenumber{9.60}&\maintablenumber{31.55}&\maintablenumber{11.37}&\maintablenumber{18.32} &\maintablenumber{3.78}&\maintablenumber{6.18}&\maintablenumber{24.37}&\maintablenumber{8.40}&\maintablenumber{13.83}\\
\midrule
\multicolumn{16}{l}{\emph{Controlled comparisons: training objective / retrieval rule}}\\
Base / AR only & \maintablenumber{7.50} & \maintablenumber{12.69} & \maintablenumber{37.40} & \maintablenumber{11.12} & \maintablenumber{18.79} & \maintablenumber{6.03} & \maintablenumber{10.48} & \maintablenumber{34.15} & \maintablenumber{12.43} & \maintablenumber{19.88} & \maintablenumber{6.05} & \maintablenumber{9.77} & \maintablenumber{30.89} & \maintablenumber{11.77} & \maintablenumber{18.42}\\
RARS / AR only & \maintablenumber{\second{8.57}} & \maintablenumber{13.39} & \maintablenumber{\second{38.98}} & \maintablenumber{\second{12.36}} & \maintablenumber{\second{20.11}} & \maintablenumber{6.40} & \maintablenumber{\second{11.38}} & \maintablenumber{34.86} & \maintablenumber{\second{13.65}} & \maintablenumber{\second{20.84}} & \maintablenumber{\second{6.51}} & \maintablenumber{10.29} & \maintablenumber{31.59} & \maintablenumber{12.34} & \maintablenumber{\second{19.44}}\\
Base / all levels & \maintablenumber{8.05} & \maintablenumber{\second{13.63}} & \maintablenumber{38.24} & \maintablenumber{12.00} & \maintablenumber{19.38} & \maintablenumber{\second{6.47}} & \maintablenumber{10.93} & \maintablenumber{\second{35.00}} & \maintablenumber{12.91} & \maintablenumber{20.69} & \maintablenumber{6.29} & \maintablenumber{\second{10.42}} & \maintablenumber{31.16} & \maintablenumber{\second{12.69}} & \maintablenumber{18.99}\\
\rowcolor{adalignrow}\textbf{RARS / all levels} & \maintablenumber{\best{9.29}} & \maintablenumber{\best{14.36}} & \maintablenumber{\best{39.62}} & \maintablenumber{\best{13.00}} & \maintablenumber{\best{20.82}} & \maintablenumber{\best{6.87}} & \maintablenumber{\best{11.80}} & \maintablenumber{\best{35.55}} & \maintablenumber{\best{14.19}} & \maintablenumber{\best{21.53}} & \maintablenumber{\best{6.71}} & \maintablenumber{\best{10.89}} & \maintablenumber{\best{32.02}} & \maintablenumber{\best{13.23}} & \maintablenumber{\best{19.80}}\\
\bottomrule
\end{tabular}}
\captionsetup{justification=justified,singlelinecheck=false}
\caption{Retrieval results on multilingual ESCI (\%).
R@$K$ and N@$K$ denote Recall@$K$ and NDCG@$K$. Controlled comparisons use matched
backbones and report means across five seeds.}
\label{tab:main-results}
\end{table}
\subsection{Experimental Setup}
\label{sec:exp-setup}
\noindent\textbf{(1) Datasets.}
We use the category-filtered US (English), ES (Spanish), and JP (Japanese)
subsets of the ESCI product-search benchmark~\citep{reddy2022esci}.
Query--product pairs are labeled Exact, Substitute, Complement, or Irrelevant.
The first three labels receive weights $3$, $2$, and $1$,
normalized per query.
Evaluation retains the original graded judgments.
\noindent\textbf{(2) Baselines.}
Published baselines include sparse retrieval (BM25),
dense retrieval (DPR, MPNet, Sentence-T5, Sentence-mT5, BGE-M3, and LaSER),
and generative retrieval (DSI$_{\mathrm{naive}}$,
DSI$_{\mathrm{semantic}}$, TIGER, Hi-Gen, LTRGR, RIPOR, CAT-ID$^2$,
MERGE, FORGE, HGRec, and CaLIR).
Controlled comparisons evaluate matched Base and RARS models under
AR-only and all-level retrieval, fixing the backbone, data, SIDs,
and decoding settings.
\noindent\textbf{(3) Implementation Details.}
Following CaLIR~\citep{zhang2026calir}, we use T5-base for US and
mT5-base for ES and JP.
Default SIDs comprise four residual quantization levels with 256 codewords each.
We train matched models for 300 epochs using AdamW with a learning
rate of $5\times10^{-4}$ and a per-device batch size of 512.
We report five-seed means from final checkpoints.
Retrieval uses trie-constrained beam search with beam size 100 and $C=3$.
All-level scoring uses $\lambda_{\mathrm{comp}}=2$, $K_{\mathrm{lex}}=2$,
$\eta_{\mathrm{lex}}=0.25$, and $m=(1,1,1,1)$.
We report Recall@$K$ ($K=5,10,100$) and NDCG@$K$ ($K=10,100$),
with Recall@100 and NDCG@10 as the primary metrics.
Training settings are fixed beforehand, and retrieval hyperparameters
are selected on validation data and fixed before testing.
Subsequent sweeps assess sensitivity without revising these settings.
Dataset statistics, baseline protocols, and implementation and statistical
details appear in Appendices~\ref{app:reproducibility},
\ref{app:rq1-ttest}, and~\ref{app:training-controls}.

\subsection{Experimental Results and Discussion}
\label{sec:rq1}
\noindent\textbf{Retrieval Effectiveness (RQ1).}
Table~\ref{tab:main-results} compares RARS with sparse, dense, and
generative retrieval methods, including matched variants that distinguish
the contributions of training and retrieval scoring.
AR-only retrieval follows Eq.~\eqref{eq:inference-ar};
all-level retrieval adds the prefix compatibility terms in
Eq.~\eqref{eq:rars-score}.
Under AR-only retrieval, RARS improves Recall@100 and NDCG@10 over
the matched base in each locale.
Since the hierarchical predictor is discarded at inference, these gains
support the benefit of multiresolution relevance supervision under
standard autoregressive decoding.
All-level scoring further improves both training variants while preserving
this advantage.
The proposed objective improves autoregressive retrieval, while compatibility
scoring yields additional gains under the augmented retrieval rule.
These improvements remain consistent across locales and primary metrics.
The resulting configuration also compares favorably with sparse, dense,
and generative baselines. The matched comparisons use a common training
and evaluation protocol with the backbone and decoding settings fixed,
allowing the contribution of the proposed objective to be assessed directly.

\noindent\textbf{Contributions of the Training Objective (RQ2).}
Table~\ref{tab:training-controls} compares five training objectives under
the same all-level retrieval rule.
This isolates differences in training supervision.
The matched base uses full-SID supervision.
Grouped soft targets weight complete identifiers by relevance.
AR-decoder soft targets apply projected conditional distributions directly
to the decoder.
Sampled tree and RARS share the hierarchical predictor in
Eq.~\eqref{eq:tree-score}, using sampled one-hot and explicit conditional
targets, respectively.
These comparisons separate the effects of target aggregation, prediction
site, and target stochasticity.
The predictor is used only during training and discarded at inference.
The AR-decoder control shows that projected local distributions are
beneficial even when applied directly to the decoder.
The sampled-tree comparison isolates the benefit of explicit conditional
supervision within the same predictor, with distributional targets
consistently outperforming sampled one-hot targets.
Improvements over grouped soft targets indicate that supervising relevance
allocation at SID refinements through the shared query representation
provides additional benefit beyond relevance weighting over complete
identifier paths.
Paired five-seed comparisons show positive gains over grouped soft-target
supervision on both primary metrics across locales, as reported in
Appendix~\ref{app:training-controls}.

\begin{table}[!tbp]
\centering
\resizebox{\textwidth}{!}{%
\begin{tabular}{lrrrrrrrrrrrrrrr}
\toprule
& \multicolumn{5}{c}{ESCI-US} & \multicolumn{5}{c}{ESCI-ES} & \multicolumn{5}{c}{ESCI-JP}\\
\cmidrule(lr){2-6}\cmidrule(lr){7-11}\cmidrule(lr){12-16}
Training objective & R@5 & R@10 & R@100 & N@10 & N@100 & R@5 & R@10 & R@100 & N@10 & N@100 & R@5 & R@10 & R@100 & N@10 & N@100\\
\midrule
Matched base & \maintablenumber{8.05} & \maintablenumber{13.63} & \maintablenumber{38.24} & \maintablenumber{12.00} & \maintablenumber{19.38} & \maintablenumber{6.47} & \maintablenumber{10.93} & \maintablenumber{35.00} & \maintablenumber{12.91} & \maintablenumber{20.69} & \maintablenumber{6.29} & \maintablenumber{10.42} & \maintablenumber{31.16} & \maintablenumber{12.69} & \maintablenumber{18.99}\\
AR-decoder soft target & \maintablenumber{8.74} & \maintablenumber{13.93} & \maintablenumber{39.04} & \maintablenumber{12.46} & \maintablenumber{20.36} & \maintablenumber{6.68} & \maintablenumber{11.33} & \maintablenumber{35.17} & \maintablenumber{13.77} & \maintablenumber{21.06} & \maintablenumber{6.57} & \maintablenumber{10.60} & \maintablenumber{31.70} & \maintablenumber{12.88} & \maintablenumber{19.42}\\
Grouped soft-target & \maintablenumber{8.61} & \maintablenumber{13.69} & \maintablenumber{38.88} & \maintablenumber{12.28} & \maintablenumber{20.03} & \maintablenumber{6.55} & \maintablenumber{11.27} & \maintablenumber{35.08} & \maintablenumber{13.48} & \maintablenumber{20.91} & \maintablenumber{6.46} & \maintablenumber{10.43} & \maintablenumber{31.61} & \maintablenumber{12.71} & \maintablenumber{19.26}\\
Sampled tree & \maintablenumber{8.89} & \maintablenumber{14.01} & \maintablenumber{39.18} & \maintablenumber{12.55} & \maintablenumber{20.29} & \maintablenumber{6.63} & \maintablenumber{11.51} & \maintablenumber{35.28} & \maintablenumber{13.69} & \maintablenumber{21.19} & \maintablenumber{6.52} & \maintablenumber{10.55} & \maintablenumber{31.81} & \maintablenumber{12.79} & \maintablenumber{19.55}\\
\rowcolor{adalignrow}RARS & \maintablenumber{\textbf{9.29}} & \maintablenumber{\textbf{14.36}} & \maintablenumber{\textbf{39.62}} & \maintablenumber{\textbf{13.00}} & \maintablenumber{\textbf{20.82}} & \maintablenumber{\textbf{6.87}} & \maintablenumber{\textbf{11.80}} & \maintablenumber{\textbf{35.55}} & \maintablenumber{\textbf{14.19}} & \maintablenumber{\textbf{21.53}} & \maintablenumber{\textbf{6.71}} & \maintablenumber{\textbf{10.89}} & \maintablenumber{\textbf{32.02}} & \maintablenumber{\textbf{13.23}} & \maintablenumber{\textbf{19.80}}\\
\bottomrule
\end{tabular}}
\captionsetup{justification=justified,singlelinecheck=false}
\caption{Training-objective comparison across all three ESCI locales under a
common all-level retrieval rule (\%). Values are means over five training
seeds for each locale.}
\label{tab:training-controls}
\end{table}

\begin{figure}[!htbp]
\centering
\newsavebox{\relevanceevidencebox}
\newsavebox{\relevancetablebox}
\sbox{\relevancetablebox}{%
  \begin{minipage}{0.485\textwidth}
  \centering\begingroup
\scriptsize
\setlength{\tabcolsep}{3pt}
\renewcommand{\arraystretch}{0.84}
\setlength{\aboverulesep}{1.5pt}
\setlength{\belowrulesep}{1.5pt}
\setlength{\cmidrulesep}{1pt}
\resizebox{\linewidth}{!}{%
\begin{tabular}{lcccccc}
\toprule
& \multicolumn{3}{c}{Recall@100} & \multicolumn{3}{c}{NDCG@10}\\
\cmidrule(lr){2-4}\cmidrule(lr){5-7}
Condition & Base & RARS & $\Delta$ & Base & RARS & $\Delta$\\
\midrule
\multicolumn{7}{l}{\emph{(a) Identifier geometry}}\\
RQ-VAE, 4\texttimes{}256 & 38.24 & 39.62 & +1.38 & 12.00 & 13.00 & +1.00\\
Hier. $k$-means, 20\textsuperscript{3} & 36.84 & 37.65 & +0.81 & 11.21 & 11.95 & +0.74\\
RQ-VAE, 5\texttimes{}256 & 37.55 & 38.60 & +1.05 & 11.54 & 12.42 & +0.88\\
\midrule
\multicolumn{7}{l}{\emph{(b) Training positives}}\\
E only & 35.91 & 36.48 & +0.57 & 10.86 & 11.35 & +0.49\\
E + S & 37.56 & 38.44 & +0.88 & 11.48 & 12.21 & +0.73\\
E + S + C & 38.24 & 39.62 & +1.38 & 12.00 & 13.00 & +1.00\\
\bottomrule
\end{tabular}}
\par
\endgroup

  \end{minipage}}
\typeout{RARS-TABLE-HEIGHT=\the\dimexpr\ht\relevancetablebox+\dp\relevancetablebox\relax}
\sbox{\relevanceevidencebox}{%
  \includegraphics[width=0.485\textwidth,
    height=\dimexpr\ht\relevancetablebox+\dp\relevancetablebox\relax]{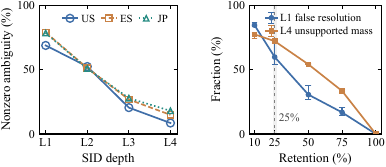}}
\begin{minipage}[t]{0.485\textwidth}
\vspace{0pt}
\begin{minipage}[c][\dimexpr\ht\relevanceevidencebox+\dp\relevanceevidencebox\relax][b]{\linewidth}
\centering
\usebox{\relevanceevidencebox}
\end{minipage}
\captionsetup{font=small,justification=justified,singlelinecheck=false,skip=5pt}
\caption{Left: ambiguity by SID depth. Right: evidence sparsity,
averaged over US, ES and JP. Bars span the minimum and maximum
across locales.}
\vspace{-0.5pt}
\label{fig:relevance-evidence}
\end{minipage}\hfill
\begin{minipage}[t]{0.485\textwidth}
\vspace{0pt}
\begin{minipage}[c][\dimexpr\ht\relevanceevidencebox+\dp\relevanceevidencebox\relax][b]{\linewidth}
\centering
\raisebox{\dp\relevancetablebox}{\usebox{\relevancetablebox}}
\end{minipage}
\captionsetup{type=table,font=small,justification=justified,singlelinecheck=false,skip=5pt}
\caption{Robustness on ESCI-US across identifier geometries
and training positives. Five-seed means (\%); $\Delta$ is the gain over
Base in percentage points.}
\vspace{-0.5pt}
\label{tab:training-robustness}
\end{minipage}
\end{figure}

\noindent\textbf{Relevance Structure and Evidence Fidelity (RQ3).}
Figure~\ref{fig:relevance-evidence} characterizes observed relevance
across SID resolutions.
Refinement ambiguity is concentrated at coarse and intermediate depths
and becomes less prevalent toward the leaves.
Across the three locales, roughly $70\%$ to $80\%$ of queries eligible
for retrieval are ambiguous at L1, about half at L2, and fewer than
$20\%$ at L4.
We assess evidence fidelity by uniformly subsampling judgments for queries
with at least two observed positives.
At $25\%$ nominal retention, more than half of these queries appear resolved
at L1 despite spanning multiple branches in the full observed set,
while roughly $70\%$ of L4 relevance mass has no retained support.
The right panel reports unweighted means across locales, with bars spanning
the minimum and maximum at each retention rate.
Incomplete judgments can make coarse refinements appear deterministic
while leaving relevance at finer depths unsupported.
RARS preserves the multiresolution structure supported by observed evidence,
so its target fidelity depends on judgment coverage.
Additional hierarchy controls and analyses of evidence subsampling appear
in Appendix~\ref{app:resolution-diagnostics}.

\noindent\textbf{Identifier Geometry and Training Relevance (RQ4).}
Table~\ref{tab:training-robustness} evaluates the robustness of RARS across
SID constructions and training relevance definitions.
We compare the default RQ-VAE hierarchy with hierarchical $k$-means and a
deeper RQ-VAE, retraining the matched base and the proposed method under the
same training protocol for each identifier space.
The method improves both primary metrics across all three constructions,
showing that the gains are not tied to a particular SID geometry.
We next vary the training positives from Exact-only to E+S and E+S+C while
keeping the evaluation protocol and test judgments fixed.
The method improves over the matched base under every relevance definition,
including Exact-only training, so the gains do not rely on treating Complement
items as positive evidence.
The larger improvements with richer relevance sets are descriptive and do
not isolate their source.
On ESCI-US, every tested nonzero tree-loss weight improves both primary
metrics over the matched base.
Using 32 high-scoring sibling negatives outperforms random negatives and
approaches full legal-sibling normalization.
Using only relevance-bearing children also improves over the matched base.
These results show robustness to hierarchy design, relevance definition,
and local candidate construction across the tested configurations.

\raggedbottom
\begin{figure}[!t]
\centering
\captionsetup{font=small,justification=rarsfulljustified,singlelinecheck=false}
\begin{minipage}[t]{0.49\linewidth}
\vspace{0pt}
\centering
\begin{minipage}[c][0.84in][b]{\linewidth}
\includegraphics[width=\linewidth]{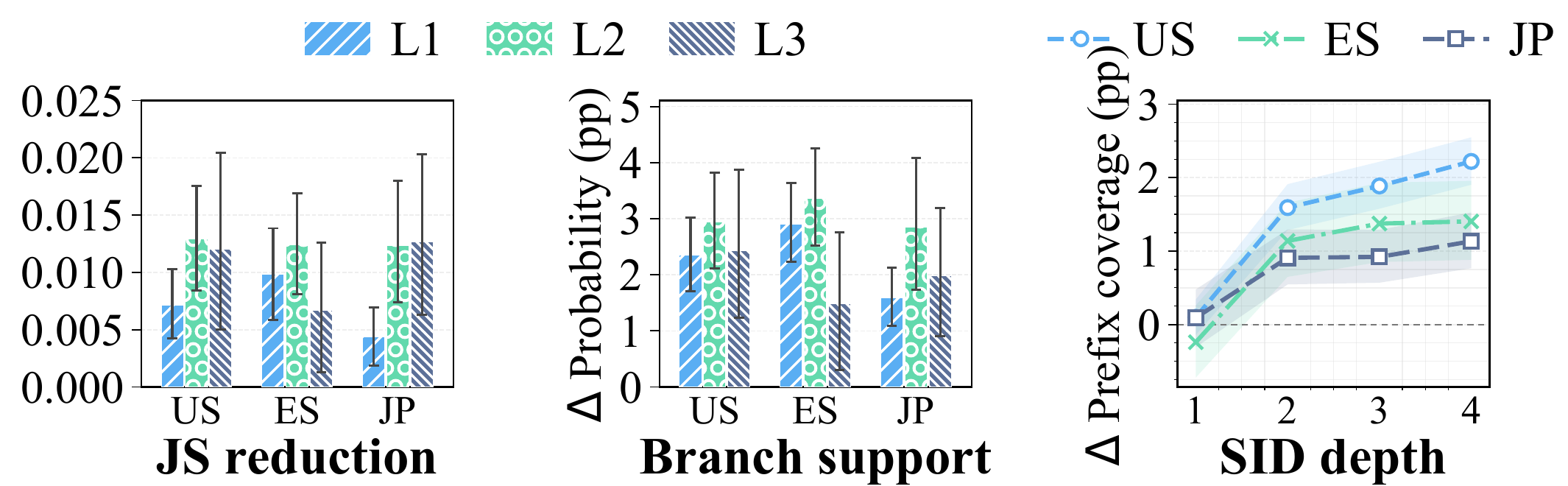}
\par\vspace{0pt}
\end{minipage}
\caption{Prediction alignment and final-ranking prefix coverage:
all-level vs.\ AR-only scoring at fixed model parameters.
Paired query-bootstrap 95\% intervals.}
\label{fig:rq3-prediction-alignment}
\end{minipage}\hfill
\begin{minipage}[t]{0.49\linewidth}
\vspace{0pt}
\centering
\begin{minipage}[c][0.84in][b]{\linewidth}
\includegraphics[width=\linewidth]{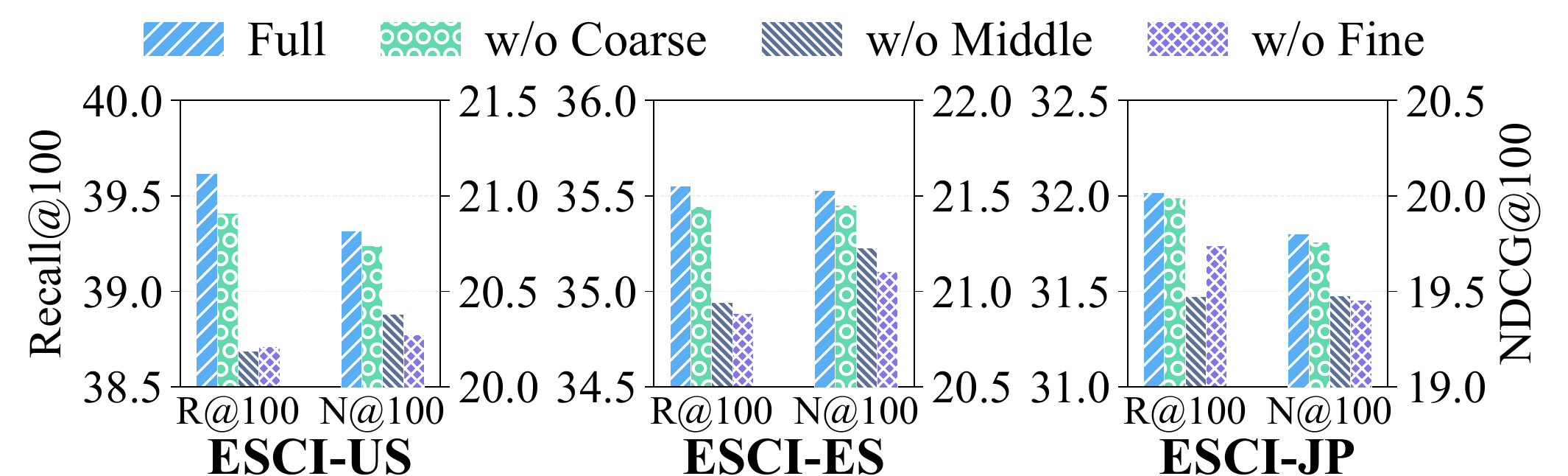}
\par\vspace{0pt}
\end{minipage}
\caption{Depth ablation with fixed model weights. Full uses all four
compatibility terms; variants remove L1, L2 or L3. Both metrics are percentages.}
\label{fig:rq4-depth-ablation}
\end{minipage}
\end{figure}

\noindent\textbf{Prediction Alignment and Prefix Coverage (RQ5).}
We examine whether compatibility scoring improves prediction at ambiguous
SID refinements. For each query-parent pair with multiple relevant children,
we compare all-level and AR-only scoring using the same model, candidate set,
retrieval budget, and normalized graded-relevance targets.
Figure~\ref{fig:rq3-prediction-alignment} shows that all-level scoring reduces
Jensen--Shannon divergence and assigns more probability mass to relevant
children across all locale--depth comparisons.
The largest improvements occur at L2, where AR-only scoring deviates most
from the relevance targets; all differences remain significant after Holm
correction.
We next assess whether these local gains persist in the final ranking.
Prefix coverage measures the fraction of relevant prefixes represented among
the top-100 retrieved SIDs.
Figure~\ref{fig:rq3-prediction-alignment} shows coverage
gains at L2--L4 in all three locales; L1 intervals include zero.
These L2--L4 gains indicate better retention of relevant branches
through successive SID refinements.
At L4, prefix coverage equals SID Recall@100 and is not a separate retrieval
endpoint; all coverage statistics are computed from final rankings.

\noindent\textbf{Contributions of Retrieval Resolutions (RQ6).}
Figure~\ref{fig:rq4-depth-ablation} examines the contribution of retrieval
resolutions by removing one compatibility term while leaving the remaining
scoring terms unchanged for the same trained model.
Across locales, the full scoring function achieves the highest Recall@100
and NDCG@100.
Removing the intermediate and fine-level terms produces the largest drops,
particularly on ESCI-US, whereas the contribution of the coarsest level is
smaller and statistically significant only on ESCI-US.
This indicates that useful retrieval evidence is distributed across multiple
stages of the SID hierarchy.
Intermediate and fine resolutions contribute complementary retrieval
information not captured by the leaf-level score alone.
Resolution-aware scoring improves both the matched base and RARS.
Using all compatibility levels increases NDCG@10 across locales under both
training conditions, showing that the benefit is not specific to
resolution-aligned training.
The AR-only comparisons further show that hierarchical supervision and
multiresolution scoring improve different parts of the retrieval process and
remain consistently effective when applied in combination.

\Needspace{24\baselineskip}
\begin{figure}[!htbp]
  \centering
  \includegraphics[width=\linewidth]{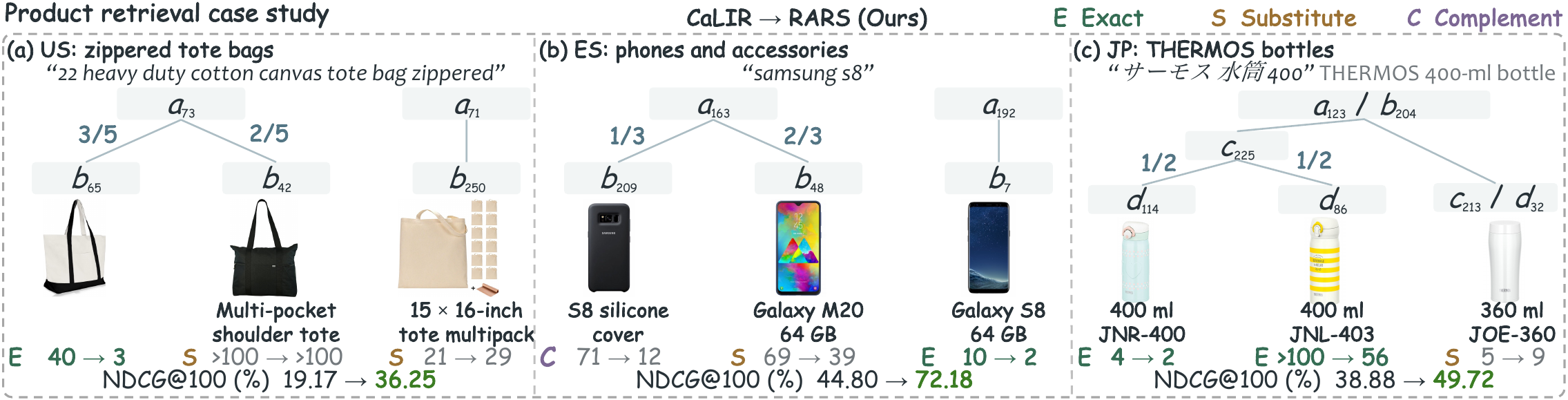}
  \captionsetup{justification=rarsfulljustified,singlelinecheck=false}
  \caption{ESCI retrieval examples. Arrows show CaLIR $\rightarrow$ RARS
  ranks and NDCG@100 (\%); $>100$ denotes absence from the top-100 results.
  Branch fractions give conditional relevance targets.}
  \label{fig:rq-case-study}
\end{figure}

\phantomsection
\noindent\textbf{Qualitative Analysis (RQ7).}
\label{sec:qualitative-analysis}%
Figure~\ref{fig:rq-case-study} shows that relevant products separate at
different depths of the SID hierarchy.
In the US example, RARS promotes the Exact zippered tote despite its shared
coarse prefix with a Substitute shoulder tote.
The ES example spans distinct coarse branches: RARS improves the rankings
of the Exact phone, a Substitute, and a Complement accessory.
In the JP example, the two Exact bottles share a prefix until the final
refinement; RARS retrieves a previously missed Exact bottle and moves the
smaller Substitute down.
These cases illustrate why relevance supervision must account for both
shared prefixes and competing refinements.
RARS learns how relevance is distributed among branches at each depth,
preserving support for multiple relevant products while distinguishing
their relevance grades.
Appendix~\ref{app:real-cases} provides additional examples.

\FloatBarrier

\Needspace{10\baselineskip}
\section{Conclusion}
\label{sec:conclusion}

RARS learns multiresolution relevance by supervising branch allocations in
the SID hierarchy. A shared query representation captures consistent
conditional targets induced by document relevance, and the auxiliary
predictor is discarded after training to preserve standard autoregressive
retrieval. Experiments on three ESCI locales show consistent gains over
matched training controls across SID constructions and relevance definitions,
including Exact-only supervision. A controlled NQ320K comparison further
demonstrates the effectiveness of retrieval scoring on Wikipedia documents,
with strong gains for documents absent from the training annotations.
RARS connects document relevance to retrieval decisions throughout the SID
hierarchy, from coarse prefixes to individual documents.

\subsection*{AI use statement}

We used generative AI tools to improve the readability and language of the
manuscript. We did not use generative AI tools
to develop the research methodology, design experiments, formulate hypotheses,
or interpret the experimental results. All AI-assisted content and code were
reviewed and verified by the authors. We take responsibility for the final
content of this work, including text, claims, and artifacts produced with the
aid of generative AI.

\subsection*{Ethics statement}

We have considered the potential ethical implications of this work and do not identify any specific ethical concerns that require further discussion.

\subsection*{Reproducibility statement}

We have made efforts to support the reproducibility of this work by providing detailed descriptions of the proposed methodology, experimental setup, and evaluation protocol in the main paper and appendix. We have also released the source code and implementation details through our GitHub repository to facilitate reproduction of our results.

\bibliography{iclr2027_conference}
\bibliographystyle{iclr2027_conference}

\clearpage
\appendix
\addtocontents{toc}{\protect\setcounter{tocdepth}{2}}

\raggedbottom
\raggedbottom
\setlength{\parskip}{4pt}
\clubpenalty=10000
\widowpenalty=10000
\displaywidowpenalty=10000
\captionsetup{font=small,skip=4pt,justification=justified,singlelinecheck=false}
\setlength{\textfloatsep}{8pt plus 2pt minus 2pt}
\setlength{\floatsep}{8pt plus 2pt minus 2pt}
\setlength{\intextsep}{8pt plus 2pt minus 2pt}
\setcounter{topnumber}{3}
\setcounter{bottomnumber}{2}
\setcounter{totalnumber}{5}
\renewcommand{\topfraction}{0.90}
\renewcommand{\bottomfraction}{0.85}
\renewcommand{\textfraction}{0.08}
\renewcommand{\floatpagefraction}{0.90}
\newcommand{\appendixtablestyle}{%
  \footnotesize
  \setlength{\tabcolsep}{4pt}%
  \renewcommand{\arraystretch}{1.04}%
}
\AddToHook{cmd/section/before}{\Needspace{5\baselineskip}}

\begingroup
\small
\hypersetup{linkcolor=black}
\renewcommand{\contentsname}{Appendix Contents}
\tableofcontents
\medskip

\endgroup

\section{Experimental Setup and Implementation}
\label{app:reproducibility}

All controlled comparisons use the same data partitions, backbone family,
and evaluation protocol. The settings below specify the reference training
and retrieval configurations.

\subsection{Datasets and evaluation metrics}
\label{app:datasets}
We use the category-filtered US, ES, and JP subsets of ESCI. Table~\ref{tab:data} gives catalog and query--product statistics. The eligible test sets contain 6,014 US, 1,656 ES, and 1,883 JP queries.

\begin{table}[!htbp]
\centering\appendixtablestyle

\begin{tabular*}{\linewidth}{@{\extracolsep{\fill}}lrrrrr@{}}
\toprule
& & \multicolumn{2}{c}{Train} & \multicolumn{2}{c}{Test}\\
\cmidrule(lr){3-4}\cmidrule(lr){5-6}
Dataset & Products & Queries & Q--P & Queries & Q--P\\
\midrule
ESCI-US & 288,372 & 20,109 & 292,354 & 6,137 & 29,971\\
ESCI-ES & 101,957 & 5,421 & 108,587 & 1,877 & 14,554\\
ESCI-JP & 119,052 & 6,543 & 127,788 & 2,163 & 15,861\\
\bottomrule
\end{tabular*}
\caption{ESCI statistics after category filtering. Q--P denotes query--product pairs.}
\label{tab:data}
\end{table}

\paragraph{Retrieval metrics.}
Let $G_q$ contain the evaluated Exact, Substitute, and Complement products for query $q$. Recall@$K$ is $|\pi_q^{1:K}\cap G_q|/|G_q|$, where $\pi_q^{1:K}$ lists the top-$K$ products. NDCG assigns gains $g(q,d)=3,2,1,0$ to Exact, Substitute, Complement, and Irrelevant:
\[
 \mathrm{DCG@}K(q)=\sum_{j=1}^{K}
 \frac{g(q,\pi_q(j))}{\log_2(j+1)},\qquad
 \mathrm{NDCG@}K(q)=\frac{\mathrm{DCG@}K(q)}{\mathrm{IDCG@}K(q)}.
\]
IDCG applies the same discount to gains in decreasing order.
We report query-averaged recall and NDCG as percentages and keep the
evaluation weights fixed across training-evidence ablations.

\subsection{Baselines}
\label{app:baselines}

Table~\ref{tab:main-results} includes sparse, dense, and generative retrieval
baselines. The matched
\textbf{Base} models are retrained under the common ESCI protocol, providing
controlled comparisons with RARS. The methods are summarized below.

\paragraph{Sparse retrieval.}
\begin{itemize}[leftmargin=*,labelindent=0pt,topsep=0pt,itemsep=1pt]
    \item \textbf{BM25}~\citep{robertson2009bm25} combines term frequency, inverse document frequency, and length normalization to score query--product relevance.
\end{itemize}

\paragraph{Dense retrieval.}
\begin{itemize}[leftmargin=*,labelindent=0pt,topsep=2pt,itemsep=2pt]
    \item \textbf{DPR}~\citep{karpukhin2020dpr} trains separate query and passage encoders with a contrastive objective. Products are ranked by the inner product of their representations with the query representation.
    \item \textbf{MPNet}~\citep{song2020mpnet} combines masked and permuted language-model pretraining. Its sentence-embedding variant represents queries and products as vectors for similarity-based retrieval.
    \item \textbf{Sentence-T5}~\citep{ni2022sentencet5} learns sentence embeddings by pooling T5 encoder states and training semantically related texts to have similar representations. The resulting embeddings support dense retrieval based on pretrained sequence-to-sequence representations.
    \item \textbf{Sentence-mT5}~\citep{yano-etal-2024-multilingual} extends Sentence-T5 with multilingual mT5 encoders and sentence-embedding data, providing a dense baseline for ESCI-ES and ESCI-JP.
    \item \textbf{BGE-M3}~\citep{chen2024bgem3} supports multilingual dense, multi-vector, and sparse retrieval, with self-knowledge distillation across these signals. We report the dense retrieval configuration.
    \item \textbf{LaSER}~\citep{jin2026laser} distills explicit reasoning into dense retrieval representations. Dual-view self-distillation aligns explicit and latent reasoning through output and intermediate-trajectory objectives, removing the need to generate a rationale at inference.
\end{itemize}

\paragraph{Generative retrieval.}
\begin{itemize}[leftmargin=*,labelindent=0pt,topsep=2pt,itemsep=2pt]
    \item \textbf{DSI$_{\mathrm{naive}}$}~\citep{tay2022dsi} directly generates non-semantic document identifiers. The retriever learns query-to-identifier associations without a semantic hierarchy over the identifiers.
    \item \textbf{DSI$_{\mathrm{semantic}}$}~\citep{tay2022dsi} assigns identifiers by hierarchical semantic clustering. Related documents share prefixes, which constrained beam search follows to retrieve documents.
    \item \textbf{TIGER}~\citep{rajput2023tiger} represents items with tuples of discrete semantic codewords. An autoregressive Transformer predicts the next item identifier from interaction context.
    \item \textbf{Hi-Gen}~\citep{wu2024higen} combines metric learning for semantic and efficiency-aware item representations with hierarchical clustering guided by product categories. A position-aware objective supervises identifier generation for personalized e-commerce search.
    \item \textbf{LTRGR}~\citep{li2024ltrgr} adds ranking-oriented training to generative retrieval. A margin-based objective over retrieved passages supplements token-level identifier likelihood.
    \item \textbf{RIPOR}~\citep{zeng2024ripor} combines relevance-oriented identifiers with progressive pairwise ranking. Its objective favors relevant over non-relevant identifiers at every prefix depth.
    \item \textbf{MERGE}~\citep{zhang2025merge} aligns documents and identifiers through multi-relevance query--document supervision. Outer-level contrastive learning models binary relevance, while inner-level learning distinguishes relevance grades and preserves identifier uniqueness.
    \item \textbf{CaLIR}~\citep{zhang2026calir} learns category-guided latent intent states before SID decoding. Hierarchical semantic reasoning and query-wise multi-positive reasoning supervise category paths, and a query-specific prefix trie restricts generation to category-consistent identifiers.
    \item \textbf{CAT-ID$^2$}~\citep{liu2026catid} incorporates the product category tree into identifier learning. Its hierarchical class constraint injects category structure during quantization, while cluster-scale and dispersion losses balance token usage and distinguish reconstructed product representations.
    \item \textbf{FORGE}~\citep{fu2026forge} examines semantic-identifier construction at industrial scale. It compares design choices and proposes efficient identifier-quality metrics associated with retrieval performance, with offline and online evaluations.
    \item \textbf{HGRec}~\citep{zhang2026hgrec} uses a hyperbolic residual vector-quantized autoencoder for item hierarchies. Differential-length codebooks adapt identifier capacity before autoregressive generation.
\end{itemize}

\paragraph{Matched training reference.}
\begin{itemize}[leftmargin=*,labelindent=0pt,topsep=2pt,itemsep=2pt]
    \item \textbf{Base} is the matched CaLIR retriever used for our controlled comparisons. Its loss consists of full-SID likelihood, category classification, and multi-positive category alignment. The ``AR only'' and ``all levels'' rows use autoregressive scoring and compatibility-augmented scoring, respectively.
\end{itemize}

\subsection{Implementation and optimization}
\label{app:implementation}
\label{app:base-objective}

\begin{table}[!htbp]
\centering\appendixtablestyle

\begin{tabularx}{\linewidth}{>{\raggedright\arraybackslash}p{0.22\linewidth}>{\raggedright\arraybackslash}p{0.33\linewidth}X}
\toprule
Component & Setting & Scope\\
\midrule
Reference architecture & CaLIR & Category-guided latent reasoning and SID decoding\\
English backbone & T5-base & ESCI-US\\
Spanish/Japanese backbone & mT5-base & ESCI-ES and ESCI-JP\\
Default SID & RQ-VAE, 4\texttimes{}256 & Alternative geometries evaluated separately\\
RQ-VAE optimizer & AdamW & 300 epochs; batch size 2048\\
GR optimizer & AdamW & Learning rate 5\texttimes{}10\textsuperscript{\textminus{}4}\\
GR batch size & 512 per device & Main training comparisons\\
GR duration & 300 epochs & Fixed training length\\
Main checkpoint & Final checkpoint at epoch 300 & Fixed training duration\\
Training repetitions & Five seeds & Main comparisons and added sensitivity experiments\\
Reference auxiliary loss & $\mathcal L_{\mathrm{cls}}+\mathcal L_{\mathrm{con}}$ & Coefficient $\lambda_{\mathrm{base}}$=0.1\\
Tree-loss weight & $\lambda_{\mathrm{tree}}$=0.5 & Fixed main setting; sensitivity reported separately\\
Tree-head rank / temperature & 64 / 1 & Query-level auxiliary predictor\\
Sibling-negative cap & 32 high-scoring legal siblings & Relevant children always included\\
Relevance mass & E/S/C weights 3/2/1 & Normalized within each query\\
Beam size & 100 & Trie-constrained complete-SID decoding\\
Category width & $C$=3 & Top three predicted category tries\\
Compatibility weight & $\lambda_{\mathrm{comp}}$=2 & All-level retrieval\\
Lexical features / weight & $K_{\mathrm{lex}}$=2, $\eta_{\mathrm{lex}}$=0.25 & All-level retrieval\\
Primary metrics & Recall@100 and NDCG@10 & NDCG gains: E/S/C/I =3/2/1/0\\
Query uncertainty & Paired bootstrap 95\% interval & Query variation at fixed model parameters\\
Significance & Paired randomization; Holm correction & Diagnostic comparisons specified in each experiment\\
\bottomrule
\end{tabularx}
\caption{Training, retrieval, and statistical settings for the ESCI experiments.}
\label{tab:config}
\end{table}

The reference retriever follows CaLIR~\citep{zhang2026calir}. Its decoder
forms category-supervised latent states before SID prediction.
Let $c_j(d)$ be the category of product $d$ at valid level $j$, and let
$r_j(c\mid q,c_{j-1}(d))$ be the category softmax with invalid children
masked. The classification loss is
\[
 \mathcal L_{\mathrm{cls}}(q,d)
 =-\sum_{j\leq J_d}\log r_j(c_j(d)\mid q,c_{j-1}(d)),
\]
where $J_d$ is the product's observed category depth. For query $q_i$, let
$P_{ij}$ be its positive categories at level $j$, $B_j$ the category
candidates represented in the batch, and $s_{ijc}$ the temperature-scaled
cosine similarity between the projected latent state and category prototype
$c$. The multi-positive term is
\[
 \mathcal L_{\mathrm{con}}
 =-\sum_{i,j:\,P_{ij}\ne\varnothing}\frac{1}{|P_{ij}|}
   \sum_{c\in P_{ij}}
   \log\frac{\exp s_{ijc}}{\sum_{c'\in B_j}\exp s_{ijc'}}.
\]
These terms correspond to the HSR and QRE objectives of CaLIR, each with
coefficient $0.1$. RARS adds supervision over the SID hierarchy through
the shared query encoder.

Table~\ref{tab:config} lists the shared settings. RQ-VAE is trained with
AdamW for 300 epochs at batch size 2048. The generative retriever uses
AdamW with learning rate $5\times10^{-4}$, per-device batch size 512,
and 300 epochs. Main and sensitivity comparisons use five training seeds
and evaluate the final epoch-300 checkpoints.
The tree-loss weight $\lambda_{\mathrm{tree}}=0.5$ was fixed before training.
Retrieval parameters $C=3$, $\lambda_{\mathrm{comp}}=2$, $K_{\mathrm{lex}}=2$,
and $\eta_{\mathrm{lex}}=0.25$ were selected exclusively on validation data
and fixed before test evaluation. Sensitivity analyses vary individual
components around this reference configuration.

Table~\ref{tab:main-results} reports five-seed means, with training-run
standard deviations in Appendix~\ref{app:training-controls}. Diagnostic
comparisons estimate query-level uncertainty by paired bootstrap and
randomization tests, conditional on fixed trained models. Scoring variants
share the query set, relevance judgments, SID assignments, category trie,
beam size, category width, and single-pass constrained decoding procedure.
Outcomes are paired by query within each locale.

\subsection{Sparse target construction}
\label{app:algorithm}

For each query, we merge duplicate query--document pairs and normalize the
relevance masses. Summing each leaf's mass at every ancestor gives the
nonzero projected targets. At each relevance-bearing parent, the candidate
set includes all relevant children and the selected high-scoring sibling
negatives. A shared query representation at each depth predicts the local
child distributions, which contribute to the parent-weighted objective in
Eq.~\eqref{eq:tree-loss}. A parent with one relevant child has zero target
ambiguity. Sibling negatives supply competing alternatives in its local
normalization.

Given the candidate sets, auxiliary prediction has cost linear in the number of active query-tree edges. The cost of hard-negative selection depends on the number of legal siblings scored at each relevance-bearing parent. The resolution head shares the query encoder and is discarded after training.

\subsection{Retrieval and compatibility scoring}
\label{app:retrieval-details}
Top-three category predictions define the legal SID set.
Equations~\eqref{eq:inference-ar} and~\eqref{eq:rars-score} score these
admitted sequences. For a given model, AR-only and compatibility-augmented
retrieval share the legal SID set. Matched models share category width and
trie construction but may predict different categories.

Queries and catalog text are normalized by Unicode NFKC and case folding.
The compatibility function uses color aliases, brands, numbers, and
normalized measurements. Let $\mathcal A(q)$ be the extracted attribute
constraints and $I_a(d)$ indicate a catalog match. The structured component is
\[
 f_{\mathrm{attr}}(q,d)=
 \frac{1}{|\mathcal A(q)|}
 \sum_{a\in\mathcal A(q)}
 \bigl[(1+\gamma_a)I_a(d)-\gamma_a\bigr],
\]
where $\gamma_a=0.25$ for brands and $0.20$ for other attributes; the
component is zero if no attribute is extracted. Attribute constraints
encode affirmative mentions in the query. From the query's word and CJK character features, we retain at
most two features in decreasing IDF order, each occurring in at least two
and at most 2\% of catalog items. We use
$\mathrm{IDF}(k)=\log[(|\mathcal D|+1)/(\mathrm{df}(k)+1)]$.
For this set $\mathcal K(q)$,
\[
 f_{\mathrm{lex}}(q,d)=
 \frac{\sum_{k\in\mathcal K(q)}\mathrm{IDF}(k)I_k(d)}
      {\sum_{k\in\mathcal K(q)}\mathrm{IDF}(k)},
 \qquad
 f(q,d)=f_{\mathrm{attr}}(q,d)+0.25f_{\mathrm{lex}}(q,d),
\]
with zero lexical contribution when the denominator is zero.
Prefix compatibility takes the maximum of $f(q,d)$ across items sharing
that SID prefix.
Exact computation scores catalog items and propagates their maximum scores
from the leaves to the root. For $N$ items, SID depth $L$, and $T$ trie
nodes, this requires $O(N(|\mathcal A(q)|+|\mathcal K(q)|)+T)$ work per
query and up to $O(NL)$ stored item-prefix associations. Catalog text
normalization and feature postings are query-independent, and prefix
maxima depend on the query. Feature postings provide an index that reduces
the number of candidate items evaluated. Exact prefix compatibility maximizes the joint
item score over each subtree after combining attribute and lexical evidence.
Compatibility values are added to legal token scores
with weight $\lambda_{\mathrm{comp}}=2$ at all four identifier depths.
The decoder generates up to 100 complete identifiers with beam 100;
duplicate identifiers are removed while retaining their first occurrence.
For a trained retriever, inference uses the query and catalog. Test relevance
judgments define evaluation targets and diagnostic strata.

\section{Proofs and Derivations}
\label{app:proofs}

\subsection{Hierarchical consistency}

The child subtrees of $u\in\mathcal P_{\ell-1}$ partition its descendant leaves. Hence
\begin{align}
\sum_{v\in\Children(u)}\mu_\ell(v\mid q)
&=\sum_{v\in\Children(u)}
  \sum_{d:\pi_\ell(d)=v}\mu(d\mid q)\\
&=\sum_{d:\pi_{\ell-1}(d)=u}\mu(d\mid q)
=\mu_{\ell-1}(u\mid q).
\end{align}
Projecting leaves directly to depth $j$ equals successive marginalization from $L$ to $j$.

\subsection{Target-induced gradient variance}

For a target $Y\sim\bm\mu$, the logit gradient is $\bm g_Y=\bm p-\bm e_Y$. Since $\mathbb E[\bm e_Y]=\bm\mu$, we have $\mathbb E[\bm g_Y]=\bm p-\bm\mu=\bm g_\mu$. The centered gradient is $\bm g_Y-\bm g_\mu=\bm\mu-\bm e_Y$, and hence
\begin{align}
\mathbb E\|\bm g_Y-\bm g_\mu\|_2^2
&=\mathbb E\|\bm e_Y\|_2^2
-2\bm\mu^\top\mathbb E[\bm e_Y]
+\|\bm\mu\|_2^2\\
&=1-2\|\bm\mu\|_2^2+\|\bm\mu\|_2^2
=1-\|\bm\mu\|_2^2.
\end{align}
At fixed local predictions, one-hot target sampling induces the logit-gradient variance above.

\subsection{Entropy and target-sampling variance}
For a categorical target $\bm\mu$, Jensen's inequality gives
$-\sum_i\mu_i\log\mu_i\geq-\log\sum_i\mu_i^2$.
Combining this entropy bound with the scalar inequality $-\log x\geq1-x$ yields
\[
 1-\|\bm\mu\|_2^2\leq-\log\|\bm\mu\|_2^2\leq H(\bm\mu).
\]
Weighting by each parent's relevance mass and summing proves
Eq.~\eqref{eq:ambiguity-bound}. The bound uses entropy in nats. Applying it to the normalized empirical
entropies requires the corresponding scale factor. The target-sampling
variance is the Gini impurity, bounded above by Shannon entropy.

\subsection{Cross-entropy decomposition}

For target distribution $\mu_\ell$ and prediction $p_\ell$ with matching support,
\begin{equation}
\CE(\mu_\ell,p_\ell)=H(\mu_\ell)+\KL(\mu_\ell\|p_\ell).
\end{equation}
The target entropy is constant with respect to the model. Cross-entropy and the KL divergence in Eq.~\eqref{eq:tree-loss} have the same minimizer, $p_\ell=\mu_\ell$, on the target support.

\subsection{Full-path consistency and grouped supervision}
\label{app:path-consistency}

When each local distribution is normalized over all legal children, the
conditionals define the complete-path distribution
$p_{\mathrm{tree}}(s(d)\mid q)
=\prod_{\ell=1}^{L}
p_\ell(s_\ell(d)\mid \pi_{\ell-1}(d),q)$.
The chain rule of KL divergence yields
\begin{equation}
D_{\mathrm{KL}}\!\left(
    \mu_L(\cdot\mid q)
    \,\middle\|\,
    p_{\mathrm{tree}}(\cdot\mid q)
\right)
=
\sum_{\ell=1}^{L}
\sum_{u\in\mathcal P_{\ell-1}}
\mu_{\ell-1}(u\mid q)\,
D_{\mathrm{KL}}\!\left(
    \mu_\ell(\cdot\mid u,q)
    \,\middle\|\,
    p_\ell(\cdot\mid u,q)
\right).
\label{eq:global-kl}
\end{equation}
Parent-mass weighting recovers full-path matching from the local objectives.

Grouped full-SID supervision optimizes
$\mathcal L_{\mathrm{group}}(q)
=-\sum_d\mu(d\mid q)\log p_\theta(s(d)\mid q)$.
With matched path parameterization $p_\theta=p_{\mathrm{tree}}$ and
full-child normalization, Eq.~\eqref{eq:global-kl} gives
\[
\mathcal L_{\mathrm{group}}(q)=\mathcal L_{\mathrm{tree}}(q)
+H(\mu_L(\cdot\mid q)).
\]
RARS uses a query-level predictor and candidate sets containing all
relevance-bearing children and selected legal sibling negatives.
Restricted candidate sets define a local discriminative objective.
Equation~\eqref{eq:global-kl} applies under full legal-child normalization.

A set-mass loss
$-\log\sum_{d\in\mathcal D_q^+}p_\theta(s(d)\mid q)$
constrains the total probability of relevant documents. RARS specifies
how this mass is allocated among children at each refinement.

\section{Supplementary Experimental Results}
\label{app:planned}

We report supplementary NQ320K retrieval results and examine training
objectives, SID and supervision choices, local prediction alignment, and
retrieval scoring. Training-run variability is reported separately from
query-level uncertainty.

\subsection{Retrieval effectiveness on NQ320K}
\label{app:nq320k}

\paragraph{Dataset and evaluation.}
We evaluate RARS's retrieval scoring on NQ320K, extending the evaluation
from product search to Wikipedia retrieval.
The benchmark contains 109,739 Wikipedia documents and 307,373 annotated
training pairs, with one relevant document per query.
Following \citet{sun2023tokenize}, we divide the 7,830 test queries into
6,075 seen queries and 1,755 unseen queries according to whether their
relevant documents appear in the training annotations.
Both subsets are evaluated over the same document collection.
The full test set measures overall retrieval effectiveness, while the
partitioned results distinguish performance by document exposure during
training.

\paragraph{Training and retrieval.}
We compare RARS with TIGER, MERGE, and CaLIR under a common evaluation
protocol. The baselines use T5-base, with all query--document supervision
drawn from the original annotations. Training runs for 100 epochs with
AdamW, a learning rate of $5\times10^{-4}$, and a batch size of 512 per
device. We evaluate the final models using constrained beam search with
a beam size of 100. CaLIR and RARS share trained parameters, SID
constraints, and beam size, isolating the contribution of retrieval scoring.
Recall@10 and Recall@100 measure retrieval coverage at two ranking depths;
MRR@100 measures how early the relevant document appears.

\begin{table}[!htbp]
\centering\appendixtablestyle
\setlength{\tabcolsep}{2.5pt}
\begin{tabular*}{\linewidth}{@{\extracolsep{\fill}}l*{9}{r}@{}}
\toprule
& \multicolumn{3}{c}{\textbf{Full test}}
& \multicolumn{3}{c}{\textbf{Seen test}}
& \multicolumn{3}{c}{\textbf{Unseen test}}\\
\cmidrule(lr){2-4}\cmidrule(lr){5-7}\cmidrule(lr){8-10}
\textbf{Method} & R@10 & R@100 & MRR
& R@10 & R@100 & MRR
& R@10 & R@100 & MRR\\
\midrule
TIGER & 53.72 & 60.20 & 47.22
& 68.86 & 74.65 & 60.71
& 1.31 & 10.20 & 0.54\\
MERGE & 54.23 & 61.28 & 47.07
& 69.30 & 75.28 & 60.44
& \second{2.05} & 12.82 & \second{0.77}\\
CaLIR & \second{56.54} & \second{63.75} & \second{51.51}
& \second{72.72} & \second{76.21} & \second{66.24}
& 0.51 & \second{20.63} & 0.51\\
\textbf{RARS} & \best{61.19} & \best{66.25} & \best{53.30}
& \best{74.09} & \best{76.51} & \best{67.03}
& \best{16.52} & \best{30.71} & \best{5.77}\\
\midrule
\emph{Improv.} ($\Delta$) & +4.65 & +2.50 & +1.79
& +1.37 & +0.30 & +0.79
& +14.47 & +10.08 & +5.00\\
\bottomrule
\end{tabular*}
\caption{Retrieval performance on NQ320K (\%). R@$K$/MRR denote
Recall@$K$/MRR@100. CaLIR and RARS share trained parameters and SID
constraints. Bold and underlining mark the best and second-best scores;
$\Delta$ is the gain over the strongest baseline in each column
(percentage points).}
\label{tab:nq320k-results}
\end{table}
\FloatBarrier

\paragraph{Retrieval effectiveness.}
RARS achieves the strongest retrieval performance across the full, seen,
and unseen test sets (Table~\ref{tab:nq320k-results}), leading all three
baselines on every reported metric.
On the full test set, its gains over the strongest baseline for each metric
are 4.65 points in Recall@10, 2.50 in Recall@100, and 1.79 in MRR@100.
Seen queries also benefit across all three metrics, with Recall@10 reaching
74.09 and MRR@100 reaching 67.03.
The largest Recall@10 gain occurs on the unseen test subset.
RARS retrieves the relevant document within the top ten results for
16.52\% of unseen queries, achieving $8.06\times$ the strongest baseline's
Recall@10 of 2.05\%. Its Recall@100 reaches 30.71,
exceeding the strongest baseline by 10.08 points, while MRR@100 rises
from 0.77 to 5.77. The improvements span both retrieval coverage and
reciprocal rank, reaching relevant documents beyond the training
annotations. The controlled comparison with CaLIR establishes retrieval
scoring as an effective source of these gains on NQ320K.

\subsection{Retrieval effectiveness and statistical uncertainty}
\label{app:rq1-ttest}

Table~\ref{tab:main-results} reports five-seed means, and
Table~\ref{app:training-full} gives sample standard deviations for all five
metrics. Figure~\ref{fig:training-paired} summarizes within-seed differences
against grouped soft-target supervision.
Appendix~\ref{app:training-retrieval} evaluates Base and RARS under AR-only
and all-level scoring.

Training-run statistics quantify variation across seeds under the fixed
training protocol. Diagnostic bootstrap intervals and randomization tests
quantify query-level variation conditional on the evaluated models.
Diagnostic checkpoints and scoring weights were selected through exploratory
analysis; the reported intervals condition on these choices. The five-seed
comparisons use training duration and parameters fixed before training.
Retrieval significance tests compare matched conditions using paired
predictions for the same queries.

\subsection{Training objectives and retrieval controls}
\label{app:training-controls}

\subsubsection{Training objectives and variation across runs}
We compare five training objectives under the common all-level retrieval
rule in Eq.~\eqref{eq:rars-score}. The matched base uses full-SID
supervision. Grouped soft targets weight complete identifiers by their
query-conditioned relevance mass. AR-decoder soft targets apply the projected
local distributions directly to the autoregressive decoder. Sampled tree
supervision and RARS use the same auxiliary predictor, trained with
sampled one-hot targets and explicit local distributions, respectively.

\begin{figure}[!htbp]
\centering
\includegraphics[width=\linewidth]{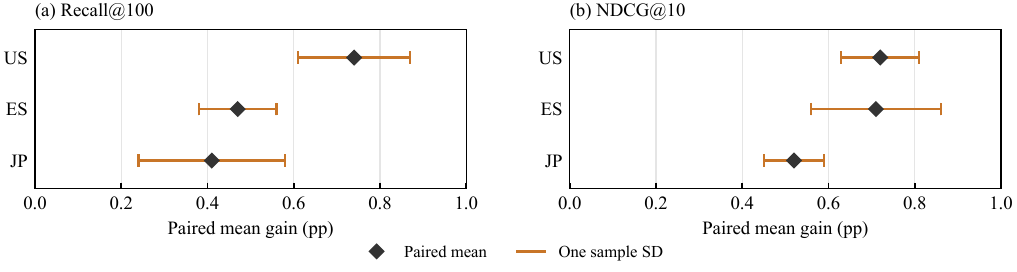}
\caption{RARS minus grouped soft-target supervision. Diamonds show the
mean of five within-seed differences; horizontal bars show one sample SD.}
\label{fig:training-paired}
\end{figure}

Table~\ref{app:training-full} reports means and sample standard deviations
over five runs for all objectives and metrics. Figure~\ref{fig:training-paired}
summarizes paired gains over grouped soft-target supervision. The Base and
RARS means correspond to the all-level results in Table~\ref{tab:main-results}.

\begin{table}[!bp]
\centering\appendixtablestyle
\begin{tabular*}{\linewidth}{@{\extracolsep{\fill}}lrrrrr@{}}
\toprule
Training condition & R@5 & R@10 & R@100 & N@10 & N@100\\
\midrule
\multicolumn{6}{l}{\textit{ESCI-US}}\\

Matched base
& 8.05\,\textpm{}\,0.08
& 13.63\,\textpm{}\,0.13
& 38.24\,\textpm{}\,0.15
& 12.00\,\textpm{}\,0.17
& 19.38\,\textpm{}\,0.11
\\

AR-decoder soft target
& 8.74\,\textpm{}\,0.06
& 13.93\,\textpm{}\,0.12
& 39.04\,\textpm{}\,0.21
& 12.46\,\textpm{}\,0.15
& 20.36\,\textpm{}\,0.13
\\

Grouped soft-target
& 8.61\,\textpm{}\,0.09
& 13.69\,\textpm{}\,0.14
& 38.88\,\textpm{}\,0.16
& 12.28\,\textpm{}\,0.08
& 20.03\,\textpm{}\,0.20
\\

Sampled tree
& 8.89\,\textpm{}\,0.11
& 14.01\,\textpm{}\,0.07
& 39.18\,\textpm{}\,0.14
& 12.55\,\textpm{}\,0.12
& 20.29\,\textpm{}\,0.18
\\

RARS
& \textbf{9.29}\,\textpm{}\,0.10
& \textbf{14.36}\,\textpm{}\,0.11
& \textbf{39.62}\,\textpm{}\,0.15
& \textbf{13.00}\,\textpm{}\,0.07
& \textbf{20.82}\,\textpm{}\,0.20
\\

\midrule

\multicolumn{6}{l}{\textit{ESCI-ES}}\\

Matched base
& 6.47\,\textpm{}\,0.07
& 10.93\,\textpm{}\,0.07
& 35.00\,\textpm{}\,0.09
& 12.91\,\textpm{}\,0.13
& 20.69\,\textpm{}\,0.09
\\

AR-decoder soft target
& 6.68\,\textpm{}\,0.08
& 11.33\,\textpm{}\,0.05
& 35.17\,\textpm{}\,0.11
& 13.77\,\textpm{}\,0.16
& 21.06\,\textpm{}\,0.07
\\

Grouped soft-target
& 6.55\,\textpm{}\,0.05
& 11.27\,\textpm{}\,0.09
& 35.08\,\textpm{}\,0.14
& 13.48\,\textpm{}\,0.10
& 20.91\,\textpm{}\,0.12
\\

Sampled tree
& 6.63\,\textpm{}\,0.06
& 11.51\,\textpm{}\,0.10
& 35.28\,\textpm{}\,0.18
& 13.69\,\textpm{}\,0.06
& 21.19\,\textpm{}\,0.15
\\

RARS
& \textbf{6.87}\,\textpm{}\,0.05
& \textbf{11.80}\,\textpm{}\,0.04
& \textbf{35.55}\,\textpm{}\,0.11
& \textbf{14.19}\,\textpm{}\,0.13
& \textbf{21.53}\,\textpm{}\,0.11
\\

\midrule

\multicolumn{6}{l}{\textit{ESCI-JP}}\\

Matched base
& 6.29\,\textpm{}\,0.10
& 10.42\,\textpm{}\,0.09
& 31.16\,\textpm{}\,0.19
& 12.69\,\textpm{}\,0.11
& 18.99\,\textpm{}\,0.13
\\

AR-decoder soft target
& 6.57\,\textpm{}\,0.08
& 10.60\,\textpm{}\,0.13
& 31.70\,\textpm{}\,0.11
& 12.88\,\textpm{}\,0.05
& 19.42\,\textpm{}\,0.17
\\

Grouped soft-target
& 6.46\,\textpm{}\,0.12
& 10.43\,\textpm{}\,0.07
& 31.61\,\textpm{}\,0.15
& 12.71\,\textpm{}\,0.09
& 19.26\,\textpm{}\,0.14
\\

Sampled tree
& 6.52\,\textpm{}\,0.07
& 10.55\,\textpm{}\,0.06
& 31.81\,\textpm{}\,0.20
& 12.79\,\textpm{}\,0.14
& 19.55\,\textpm{}\,0.09
\\

RARS
& \textbf{6.71}\,\textpm{}\,0.05
& \textbf{10.89}\,\textpm{}\,0.09
& \textbf{32.02}\,\textpm{}\,0.09
& \textbf{13.23}\,\textpm{}\,0.06
& \textbf{19.80}\,\textpm{}\,0.10
\\

\bottomrule
\end{tabular*}
\caption{Training objectives under all-level retrieval on ESCI (\%).
Values are mean $\pm$ sample SD over five runs. AR-decoder soft target is a
training objective; AR only in Table~\ref{tab:main-results} is a retrieval
setting. Retrieval follows Eq.~\eqref{eq:rars-score}.}
\label{app:training-full}
\end{table}

\paragraph{Local targets on the autoregressive decoder.}
Let $z_\theta(c\mid q,u)$ denote the decoder logit for child token $c$ after
prefix $u$. AR-decoder soft-target supervision defines
\[
 p^{\mathrm{AR}}_\theta(c\mid q,u)
 =\frac{\exp z_\theta(c\mid q,u)}
        {\sum_{c'\in\mathrm{Ch}(u)}\exp z_\theta(c'\mid q,u)},
\]
and minimizes the parent-weighted conditional KL in
Eq.~\eqref{eq:tree-loss} in addition to the reference losses.
This control applies the targets directly to the decoder.
Grouped supervision averages complete-SID negative log-likelihoods using
weights $\mu(d\mid q)$. Appendix~\ref{app:path-consistency} gives the
formal relation between grouped and conditional objectives. The controls assess
where the targets are applied: to complete SID likelihoods, to the decoder's
local predictions, or to the shared query representation through the RARS
predictor. The decoder normalizes over all legal children; RARS uses
Eq.~\eqref{eq:tree-posterior}.

\paragraph{Paired differences.}
RARS improves the five-run means of both primary metrics over the matched
base, grouped soft targets, and sampled tree
(Table~\ref{app:training-full}).
Mean NDCG@10 gains over grouped soft targets are $0.72$, $0.71$, and $0.52$
percentage points on US, ES, and JP; the corresponding Recall@100 gains
are $0.74$, $0.47$, and $0.41$ points.
Figure~\ref{fig:training-paired} summarizes these paired comparisons.
Diamonds show mean gains, and horizontal bars show one sample SD of
the five within-seed differences for each locale and metric.

\FloatBarrier
\Needspace{6\baselineskip}
\subsubsection{Training and retrieval resolution}
\label{app:training-retrieval}
We evaluate both training conditions under four retrieval settings.
The masks $0000$, $0001$, $1110$, and $1111$ enable no compatibility terms,
the leaf term only, the intermediate terms only, and all four terms,
respectively. The auxiliary prediction head is absent throughout retrieval.
Comparisons within a trained model measure the contribution of scoring;
comparisons between training conditions at the same mask measure the training
contrast under a common retrieval rule.
\begin{table}[!htbp]
\centering\appendixtablestyle
\begin{tabular*}{\linewidth}{@{\extracolsep{\fill}}lrrrrr@{}}
\toprule
Retrieval rule & Mask & Base R@100 & Base N@10 & RARS R@100 & RARS N@10\\
\midrule
\multicolumn{6}{l}{\textit{ESCI-US}}\\
AR only & 0000 & 37.40 & 11.12 & 38.98 & 12.36\\
Leaf only & 0001 & 37.76 & 11.39 & 38.87 & 12.43\\
Intermediate only & 1110 & 38.08 & 11.58 & 39.21 & 12.79\\
All levels & 1111 & 38.24 & 12.00 & 39.62 & 13.00\\
\midrule
\multicolumn{6}{l}{\textit{ESCI-ES}}\\
AR only & 0000 & 34.15 & 12.43 & 34.86 & 13.65\\
Leaf only & 0001 & 34.33 & 12.71 & 35.03 & 13.58\\
Intermediate only & 1110 & 34.55 & 12.92 & 35.32 & 13.88\\
All levels & 1111 & 35.00 & 12.91 & 35.55 & 14.19\\
\midrule
\multicolumn{6}{l}{\textit{ESCI-JP}}\\
AR only & 0000 & 30.89 & 11.77 & 31.59 & 12.34\\
Leaf only & 0001 & 31.03 & 11.96 & 31.61 & 12.69\\
Intermediate only & 1110 & 31.27 & 12.17 & 31.85 & 13.02\\
All levels & 1111 & 31.16 & 12.69 & 32.02 & 13.23\\
\bottomrule
\end{tabular*}
\caption{Training objectives evaluated under four retrieval settings (\%).
Masks indicate active compatibility terms at L1--L4; autoregressive scores
are always included. All-level results agree with
Table~\ref{app:training-full}, and the AR-only and all-level results also
appear in Table~\ref{tab:main-results}.}
\label{app:training-factorial}
\end{table}

Under AR-only retrieval, RARS improves Recall@100 by $1.58/0.71/0.70$
points and NDCG@10 by $1.24/1.22/0.57$ points on US/ES/JP.
All-level scoring adds $0.88/0.48/0.92$ NDCG@10 points for Base and
$0.64/0.54/0.89$ for RARS. The differences in these increments are
$-0.24/0.06/{-0.03}$ points. Training and compatibility scoring both improve
retrieval under matched conditions.

For a metric evaluated under AR-only and all-level retrieval, let
$B_{\mathrm{AR}}$ denote the base-training score and let
$L_{\mathrm{AR}}$ and $L_F$ denote the corresponding RARS scores.
The total gain is
\begin{equation}
L_F-B_{\mathrm{AR}}
=\underbrace{L_{\mathrm{AR}}-B_{\mathrm{AR}}}_{\text{training contrast under AR}}
+\underbrace{L_F-L_{\mathrm{AR}}}_{\text{retrieval increment within RARS}}.
\label{eq:training-retrieval-decomposition}
\end{equation}
The decomposition follows an ordered contrast: changing the training
objective under AR-only retrieval, then adding compatibility scoring to
the RARS model.

\subsection{Identifier and training sensitivity}
\label{app:training-robustness}

We vary SID construction, training positives, tree-loss weight, temperature,
sibling candidates, and head dimension. Each comparison changes one component
of the fixed reference configuration.

\subsubsection{Identifier geometry and relevance definitions}
\paragraph{Alternative identifier geometries.}
We evaluate two alternative SID constructions on ESCI-US. Hierarchical
$k$-means recursively partitions the product representations into a
three-level tree with branching factor 20 at each level, following the
CaLIR construction~\citep{zhang2026calir}. Each product receives the
sequence of cluster indices along its path through this
$20\times20\times20$ hierarchy. When products share a terminal cluster,
we append a deterministic suffix solely to ensure identifier uniqueness.
The hierarchy contains three semantic clustering levels.
The alternative RQ-VAE uses five residual codebooks with 256 entries each
($5\times256$), compared with the default four-codebook construction
($4\times256$). The two variants use the same RQ-VAE settings at different
residual depths. For each geometry, both Base and RARS are trained from scratch with the
corresponding identifiers, using the same data partitions, optimization
settings, and five random seeds.

\paragraph{Training-evidence sensitivity.}
The E-only and E+S variants use different sets of training positives
under the same evaluation protocol. E-only treats Exact-labeled products as positives, while E+S uses
Exact and Substitute products; the full setting includes Exact,
Substitute, and Complement.
The original $3/2/1$ relevance weights are retained and renormalized over the
positives available for each query. All variants retain the same test queries, product corpus, semantic
identifiers, retrieval settings, and ESCI relevance judgments. The evaluation target is fixed across training variants.

\begin{table}[!htbp]
\centering\appendixtablestyle
\begin{tabular*}{\linewidth}{@{\extracolsep{\fill}}llrrrrr@{}}
\toprule
SID geometry & Training & R@5 & R@10 & R@100 & N@10 & N@100\\
\midrule
RQ-VAE, 4\texttimes{}256 & Base & 8.05 & 13.63 & 38.24 & 12.00 & 19.38\\
 & RARS & \textbf{9.29} & \textbf{14.36} & \textbf{39.62} & \textbf{13.00} & \textbf{20.82}\\
Hier. $k$-means, 20\textsuperscript{3} & Base & 7.81 & 12.88 & 36.84 & 11.21 & 18.83\\
 & RARS & \textbf{8.36} & \textbf{13.57} & \textbf{37.65} & \textbf{11.95} & \textbf{19.54}\\
RQ-VAE, 5\texttimes{}256 & Base & 8.02 & 13.11 & 37.55 & 11.54 & 19.17\\
 & RARS & \textbf{8.84} & \textbf{13.85} & \textbf{38.60} & \textbf{12.42} & \textbf{20.13}\\
\bottomrule
\end{tabular*}
\caption{Retrieval under three SID constructions on ESCI-US. Base and
RARS are trained from scratch for each construction. The $20^3$ hierarchy
uses three clustering levels plus an identifier-uniqueness suffix; the
RQ-VAE variants differ only in residual depth. Values are five-seed means
(\%), with the better value in each pair in bold.}
\label{tab:geometry-full}
\end{table}

\begin{table}[!htbp]
\centering\appendixtablestyle
\begin{tabular*}{\linewidth}{@{\extracolsep{\fill}}llrrrrr@{}}
\toprule
Training positives & Training & R@5 & R@10 & R@100 & N@10 & N@100\\
\midrule
E only & Base & 7.52 & 12.31 & 35.91 & 10.86 & 18.35\\
 & RARS & \textbf{7.90} & \textbf{12.75} & \textbf{36.48} & \textbf{11.35} & \textbf{18.86}\\
E + S & Base & 7.97 & 13.00 & 37.56 & 11.48 & 19.15\\
 & RARS & \textbf{8.64} & \textbf{13.64} & \textbf{38.44} & \textbf{12.21} & \textbf{19.94}\\
E + S + C & Base & 8.05 & 13.63 & 38.24 & 12.00 & 19.38\\
 & RARS & \textbf{9.29} & \textbf{14.36} & \textbf{39.62} & \textbf{13.00} & \textbf{20.82}\\
\bottomrule
\end{tabular*}
\caption{Sensitivity to training relevance on ESCI-US. E/S/C denote
Exact/Substitute/Complement labels. Each pair compares Base and RARS
under the same training relevance definition and fixed evaluation judgments.
Values are five-seed means (\%); bold indicates the better value in each pair.}
\label{tab:relevance-full}
\end{table}

RARS improves all five metrics for each SID construction. NDCG@10 gains
in percentage points are $0.74$ for hierarchical $k$-means, $0.88$ for
the five-level RQ-VAE, and $1.00$ for the reference RQ-VAE.

With Exact-only positives, the gains are $0.57$ points in Recall@100 and
$0.49$ points in NDCG@10. The corresponding gains are $0.88/0.73$ with
Exact and Substitute, and $1.38/1.00$ when the training evidence includes
Exact, Substitute, and Complement products.

\subsubsection{Training hyperparameters and sibling candidates}
\label{app:training-sensitivity}
\begin{table}[!htbp]
\centering\appendixtablestyle
\begin{tabular*}{\linewidth}{@{\extracolsep{\fill}}lrrrrr@{}}
\toprule
$\lambda_{\mathrm{tree}}$ & R@5 & R@10 & R@100 & N@10 & N@100\\
\midrule
0 & 8.05 & 13.63 & 38.24 & 12.00 & 19.38\\
0.10 & 8.72 & 13.81 & 39.00 & 12.51 & 20.21\\
0.25 & 9.08 & 14.13 & 39.43 & 12.85 & 20.60\\
0.50 & \textbf{9.29} & \textbf{14.36} & 39.62 & \textbf{13.00} & 20.82\\
1.00 & 9.25 & 14.31 & 39.38 & 12.76 & \textbf{20.95}\\
2.00 & 8.96 & 14.06 & \textbf{39.71} & 12.91 & 20.63\\
\bottomrule
\end{tabular*}
\caption{Sensitivity to tree-loss weight on ESCI-US (\%). The reference
weight $0.50$ was specified before training. Bold indicates the maximum
in each metric.}
\label{tab:tree-weight}
\end{table}

\begin{table}[!htbp]
\centering\appendixtablestyle
\begin{tabular*}{\linewidth}{@{\extracolsep{\fill}}lrrrrr@{}}
\toprule
Local candidate construction & R@5 & R@10 & R@100 & N@10 & N@100\\
\midrule
Relevant children only & 8.48 & 13.77 & 38.65 & 12.18 & 19.89\\
+ 32 random sibling negatives & 8.79 & 13.93 & 39.07 & 12.57 & 20.28\\
+ 8 high-scoring siblings & 9.05 & 14.17 & 39.39 & 12.83 & 20.58\\
+ 32 high-scoring siblings & \textbf{9.29} & \textbf{14.36} & \textbf{39.62} & \textbf{13.00} & \textbf{20.82}\\
All legal siblings & 9.26 & 14.34 & 39.24 & 12.85 & 20.73\\
\bottomrule
\end{tabular*}
\caption{Sibling-negative construction on ESCI-US (\%). All candidate sets include every relevance-bearing child. Random and high-scoring negatives are legal siblings. Bold indicates the best value in each metric.}
\label{tab:sibling-negatives}
\end{table}

Every tested nonzero tree weight improves both primary metrics over
$\lambda_{\mathrm{tree}}=0$. NDCG@10 changes from $13.00$ to $12.76$ between
weights $0.5$ and $1.0$, while Recall@100 changes from $39.62$ to $39.38$.
At weight $2.0$, Recall@100 is $39.71\%$ and NDCG@10 is $12.91\%$.

Relevant-child-only normalization improves NDCG@10 from the base value of
$12.00$ to $12.18$. Adding 32 random sibling negatives raises it to $12.57$;
32 high-scoring negatives reach $13.00$. This setting exceeds full
legal-sibling normalization by $0.38$ points in Recall@100 and $0.15$
points in NDCG@10. Gains are present with relevant-child normalization
and increase when high-scoring sibling negatives are added. At a parent with one
relevant child, this restricted softmax has zero local gradient. Sibling
negatives provide competing alternatives at such parents.

\paragraph{Hyperparameter sensitivity across locales.}

\begin{figure}[!htbp]
\centering
\includegraphics[width=\linewidth]{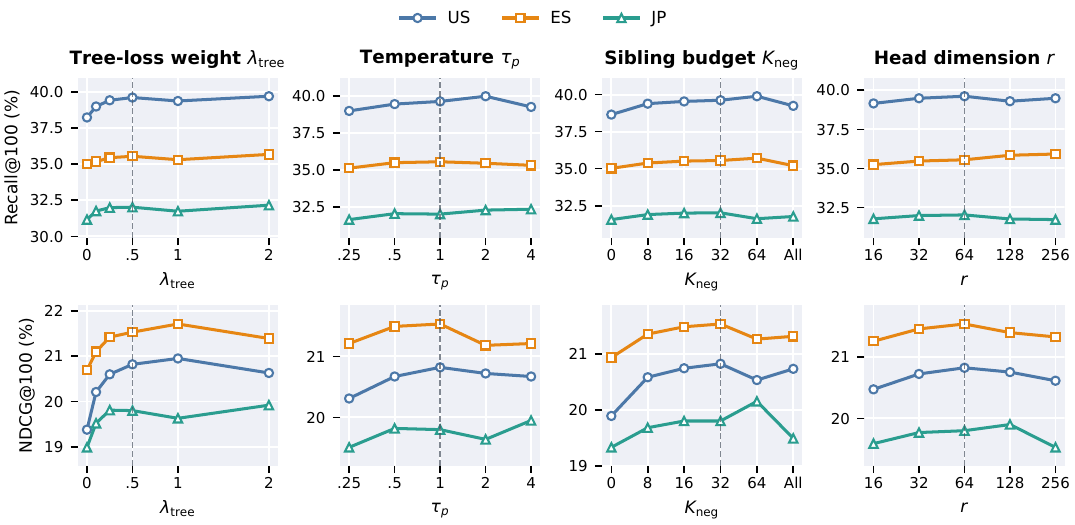}
\caption{Sensitivity to training hyperparameters under all-level retrieval
on ESCI-US, ES, and JP. The top and bottom rows report Recall@100 and
NDCG@100 (\%). Vertical dashed lines mark the reference parameter values.}
\label{fig:core-hyperparameters}
\end{figure}

Figure~\ref{fig:core-hyperparameters} evaluates the sensitivity of RARS
to its training hyperparameters under all-level retrieval. Each parameter
is varied in turn, holding the others at their reference values:
$\lambda_{\mathrm{tree}}=0.5$, $\tau_p=1$, $K_{\mathrm{neg}}=32$, and $r=64$.
Each parent's candidate set contains all relevance-bearing
children and up to $K_{\mathrm{neg}}$ high-scoring legal sibling negatives.

\paragraph{Loss weight and prediction temperature.}
Relative to $\lambda_{\mathrm{tree}}=0$, the reference weight $0.5$
improves Recall@100 and NDCG@100 in all three locales.
Recall@100 peaks at weight $2.0$ in every locale, while NDCG@100 peaks
at $1.0$ in US and ES and at $2.0$ in JP.
The temperature sweep also gives different optima: US Recall@100 peaks
at $\tau_p=2$, US and ES NDCG@100 peak at $\tau_p=1$, and both JP
metrics peak at $\tau_p=4$.

\paragraph{Sibling budget and head dimension.}
Adding eight high-scoring sibling negatives improves both metrics in
every locale. The reference budget $K_{\mathrm{neg}}=32$ exceeds full
legal-child normalization on both metrics in all three locales.
Increasing the head dimension from $r=16$ to $r=64$ also improves both
metrics in every locale. US reaches its highest values on both metrics
at $r=64$. ES Recall@100 peaks at $r=256$, while its NDCG@100 peaks
at $r=64$; JP NDCG@100 peaks at $r=128$.

\subsection{Prediction alignment and prefix coverage}
\label{app:rq3}

\paragraph{Local prediction diagnostics.}
Local prediction is evaluated on 512 held-out queries per locale using
fixed trained models. Queries with applicable constraints are stratified by
ambiguity and positive count. Of these, 457 US, 447 ES, and 451 JP queries
have at least one parent with multiple relevant children. For each such
parent, relevance grades E/S/C receive weights 3/2/1 and are normalized to a
conditional distribution over catalog-legal children. The comparison varies
compatibility scores at fixed model parameters, query, and parent prefix.

Within each query and depth, parent-level measurements are weighted by
relevance mass and normalized over the included ambiguous parents. The
figure averages queries separately at each depth, retaining all nine
locale-depth comparisons. Query counts by depth appear in Table~\ref{tab:rq3-alignment-expanded}. Confidence intervals use 10,000 paired query-bootstrap
samples. Two-sided paired sign-randomization uses 10,000 draws; Holm correction covers each metric's nine locale-depth comparisons.

\begin{table}[!htbp]
\centering\appendixtablestyle

\begin{tabular*}{\linewidth}{@{\extracolsep{\fill}}llrrrr@{}}
\toprule
\multicolumn{6}{l}{\textit{(a) Jensen--Shannon divergence}}\\
\midrule
Locale & Depth & Queries & AR-only & All-level & JS reduction [95\% CI]\\
\midrule
US & L1 & 394 & 0.3197 & 0.3126 & 0.0071 [0.0042, 0.0103]\textsuperscript{*}\\
 & L2 & 297 & 0.4531 & 0.4402 & 0.0129 [0.0084, 0.0176]\textsuperscript{*}\\
 & L3 & 160 & 0.3233 & 0.3112 & 0.0120 [0.0050, 0.0204]\textsuperscript{*}\\
\midrule
ES & L1 & 432 & 0.4054 & 0.3956 & 0.0098 [0.0059, 0.0139]\textsuperscript{*}\\
 & L2 & 280 & 0.4228 & 0.4104 & 0.0124 [0.0081, 0.0169]\textsuperscript{*}\\
 & L3 & 160 & 0.2972 & 0.2905 & 0.0067 [0.0013, 0.0126]\textsuperscript{*}\\
\midrule
JP & L1 & 431 & 0.4210 & 0.4166 & 0.0044 [0.0019, 0.0069]\textsuperscript{*}\\
 & L2 & 268 & 0.4367 & 0.4243 & 0.0124 [0.0074, 0.0180]\textsuperscript{*}\\
 & L3 & 170 & 0.2667 & 0.2540 & 0.0126 [0.0063, 0.0203]\textsuperscript{*}\\
\bottomrule
\end{tabular*}
\par\medskip
\begin{tabular*}{\linewidth}{@{\extracolsep{\fill}}llrrrr@{}}
\toprule
\multicolumn{6}{l}{\textit{(b) Probability assigned to relevant children (\%)}}\\
\midrule
Locale & Depth & Queries & AR-only & All-level & $\Delta$Support [95\% CI]\\
\midrule
US & L1 & 394 & 50.74 & 53.09 & +2.35 [1.71, 3.02]\textsuperscript{*}\\
 & L2 & 297 & 31.04 & 33.98 & +2.94 [2.11, 3.82]\textsuperscript{*}\\
 & L3 & 160 & 59.69 & 62.12 & +2.43 [1.24, 3.88]\textsuperscript{*}\\
\midrule
ES & L1 & 432 & 35.11 & 38.01 & +2.90 [2.23, 3.63]\textsuperscript{*}\\
 & L2 & 280 & 36.02 & 39.38 & +3.36 [2.52, 4.26]\textsuperscript{*}\\
 & L3 & 160 & 61.93 & 63.41 & +1.48 [0.31, 2.75]\textsuperscript{*}\\
\midrule
JP & L1 & 431 & 34.98 & 36.57 & +1.58 [1.09, 2.12]\textsuperscript{*}\\
 & L2 & 268 & 36.04 & 38.89 & +2.85 [1.73, 4.08]\textsuperscript{*}\\
 & L3 & 170 & 68.85 & 70.83 & +1.99 [0.91, 3.19]\textsuperscript{*}\\
\bottomrule
\end{tabular*}
\caption{Prediction at ambiguous refinements under AR-only and all-level
scoring, by locale and SID depth. JS measures divergence from the relevance
target; Support is the probability assigned to relevant children (\%).
Intervals are paired query-bootstrap 95\% intervals for the improvement.
$^{*}$ denotes Holm-adjusted $p<0.05$.}
\label{tab:rq3-alignment-expanded}
\end{table}

\paragraph{Ambiguity and prediction error.}
Among ambiguous parents, one standard deviation of normalized refinement
ambiguity is associated with a $0.0267$ increase in baseline JS divergence
(query-clustered 95\% CI $[0.0078,0.0456]$). The regression controls for
relevant-child count, legal-child count, depth, and locale. The coefficient
for JS reduction is $-0.0073$ (95\% CI $[-0.0120,-0.0026]$).

\paragraph{Prefix coverage on the complete evaluation sets.}
Retrieval diagnostics use one fixed trained model per locale and the full
evaluation sets, separately from the five-seed means in
Table~\ref{tab:main-results}. AR-only and all-level retrieval evaluate the
same trained model under their respective scoring rules. At each SID depth, prefix coverage is the
fraction of relevant prefixes represented among the final top-100 retrieved
SIDs. A loss event occurs when a query has positive prefix coverage at one
depth and zero at the next. Both statistics are computed from the final
top-100 rankings.

Figure~\ref{fig:rq3-prediction-alignment} reports coverage changes for
6,014 US, 1,656 ES, and 1,883 JP queries. Intervals use 10,000 paired
query-bootstrap samples and tests use 10,000 paired sign randomizations.
Holm correction covers four coverage and three loss-event comparisons
within each locale. Resampling estimates query-level variation for each
fixed model and scoring rule.

Prefix coverage improves significantly at L2--L4 in all three locales. The L4
differences are $2.22$, $1.40$, and $1.13$ percentage points on US, ES, and
JP, respectively. The corresponding L1 intervals include zero.
L4 coverage equals SID Recall@100.

\subsection{Ablation of retrieval resolutions}
\label{app:rq4}

We remove one prefix compatibility term at a time, holding model parameters
and the remaining scoring terms fixed. Training-objective comparisons are
reported in Appendix~\ref{app:training-controls}.

\begin{table}[!htbp]
\centering\appendixtablestyle

\begin{tabular*}{\linewidth}{@{\extracolsep{\fill}}llrr@{}}
\toprule
Locale & Variant & R@100 & NDCG@100\\
\midrule
\multirow{4}{*}{US}
& Full RARS & \textbf{39.6187} & \textbf{20.8169}\\
& w/o coarse (L1) & 39.4102 & 20.7389\\
& w/o middle (L2) & 38.6885 & 20.3803\\
& w/o fine (L3) & 38.7108 & 20.2723\\
\midrule
\multirow{4}{*}{ES}
& Full RARS & \textbf{35.5515} & \textbf{21.5271}\\
& w/o coarse (L1) & 35.4424 & 21.4499\\
& w/o middle (L2) & 34.9420 & 21.2282\\
& w/o fine (L3) & 34.8847 & 21.1042\\
\midrule
\multirow{4}{*}{JP}
& Full RARS & \textbf{32.0187} & \textbf{19.8000}\\
& w/o coarse (L1) & 31.9904 & 19.7607\\
& w/o middle (L2) & 31.4732 & 19.4764\\
& w/o fine (L3) & 31.7407 & 19.4557\\
\bottomrule
\end{tabular*}
\caption{Ablation of compatibility scoring by SID depth, corresponding to
Figure~\ref{fig:rq4-depth-ablation}. L1--L3 denote coarse, middle, and fine
prefix terms. All conditions use the same trained model, decoder, beam 100,
and query set, with the leaf term retained. Bold indicates the best value
within each locale.}
\label{tab:rq4-depth-full}
\end{table}

Table~\ref{tab:rq4-depth-full} shows that full scoring achieves the highest
R@100 and NDCG@100 in all three locales. Removing L2 or L3 produces the
largest decreases in both metrics. The paired differences in
Table~\ref{tab:rq4-depth-uncertainty} are statistically significant for both
resolutions across locales. L1 yields positive differences in every locale,
with significant gains on both metrics in US.
The L2 and L3 ablations retain the leaf term and all other prefix terms.
The decrease in retrieval effectiveness after removing either term shows
that both resolutions contribute to the full scoring function.
Paired tests quantify query-level variation for the fixed models and
scoring settings (Appendix~\ref{app:rq1-ttest}).

\begin{table}[!htbp]
\centering\appendixtablestyle
\begin{tabular*}{\linewidth}{@{\extracolsep{\fill}}llrcrrcr@{}}
\toprule
& & \multicolumn{3}{c}{R@100} & \multicolumn{3}{c}{NDCG@100}\\
\cmidrule(lr){3-5}\cmidrule(lr){6-8}
Locale & Term & $\Delta$ & 95\% CI & $p$ & $\Delta$ & 95\% CI & $p$\\
\midrule
\multirow{3}{*}{US}
& L1 & +0.2085 & $[0.0613,0.3602]$ & 0.0046 & +0.0780 & $[0.0247,0.1322]$ & 0.0034\\
& L2 & +0.9302 & --- & $<0.001$ & +0.4367 & --- & $<0.001$\\
& L3 & +0.9079 & --- & $<0.001$ & +0.5446 & --- & $<0.001$\\
\midrule
\multirow{3}{*}{ES}
& L1 & +0.1090 & --- & 0.4955 & +0.0772 & --- & 0.2324\\
& L2 & +0.6095 & $[0.2608,0.9867]$ & 0.0013 & +0.2990 & $[0.1607,0.4447]$ & $<0.001$\\
& L3 & +0.6668 & $[0.4093,0.9714]$ & $<0.001$ & +0.4229 & $[0.2867,0.5585]$ & $<0.001$\\
\midrule
\multirow{3}{*}{JP}
& L1 & +0.0283 & --- & 0.6955 & +0.0393 & --- & 0.3839\\
& L2 & +0.5455 & $[0.3199,0.8099]$ & $<0.001$ & +0.3236 & $[0.1923,0.4548]$ & $<0.001$\\
& L3 & +0.2780 & $[0.1361,0.4424]$ & $<0.001$ & +0.3443 & $[0.2361,0.4735]$ & $<0.001$\\
\bottomrule
\end{tabular*}
\caption{Paired differences for the retrieval-resolution ablations.
$\Delta$ is full RARS minus the ablation of the indicated term, in percentage
points. Brackets give reported 95\% paired query-bootstrap confidence
intervals; $p$-values use 10,000 two-sided paired sign randomizations.}
\label{tab:rq4-depth-uncertainty}
\end{table}

\section{Relevance Resolution and Evidence Analyses}
\label{app:resolution-diagnostics}

We analyze relevance profiles across SID hierarchies and evidence subsets.

\subsection{Relevance across identifier resolutions}
\label{sec:rq2}
\begin{table}[!htbp]
\centering
\captionsetup{justification=justified,singlelinecheck=false}
\begingroup\appendixtablestyle
\newcommand{\rqhead}[1]{#1}
\definecolor{rq2accent}{HTML}{955622}
\setlength{\tabcolsep}{2pt}
\renewcommand{\arraystretch}{1.04}
\setlength{\heavyrulewidth}{0.5pt}
\setlength{\lightrulewidth}{0.3pt}
\setlength{\cmidrulewidth}{0.25pt}
\setlength{\aboverulesep}{1pt}
\setlength{\belowrulesep}{1pt}
\setlength{\cmidrulesep}{0.7pt}
\begin{tabularx}{\linewidth}{@{}>{\raggedright\arraybackslash}Xrrrrrrrrr@{}}
\toprule
& \multicolumn{3}{c}{ESCI-US} & \multicolumn{3}{c}{ESCI-ES} & \multicolumn{3}{c}{ESCI-JP}\\
\cmidrule(lr){2-4}\cmidrule(lr){5-7}\cmidrule(lr){8-10}
\rowcolor{logicgray}
\textit{(a) Depth} & \rqhead{$\Delta_\ell$} & \rqhead{$Q$ (\%)} & \rqhead{$M$ (\%)} & \rqhead{$\Delta_\ell$} & \rqhead{$Q$ (\%)} & \rqhead{$M$ (\%)} & \rqhead{$\Delta_\ell$} & \rqhead{$Q$ (\%)} & \rqhead{$M$ (\%)}\\
\midrule
L1 & 0.700 & 68.8 & 68.8 & 1.128 & 78.9 & 78.9 & 1.127 & 78.4 & 78.4\\
L2 & 0.399 & 52.4 & 38.1 & 0.320 & 51.2 & 30.1 & 0.321 & 50.9 & 30.6\\
L3 & 0.065 & 20.5 & 8.1 & 0.071 & 26.8 & 8.2 & 0.072 & 27.9 & 8.6\\
L4 & 0.020 & 8.5 & 2.5 & 0.029 & 14.9 & 3.5 & 0.038 & 18.0 & 4.6\\
\midrule
\rowcolor{logicgray}
\textit{(b) Hierarchy} & \rqhead{$\Delta_2$} & \rqhead{$\Delta_{3:4}$} & \rqhead{$\tau$} & \rqhead{$\Delta_2$} & \rqhead{$\Delta_{3:4}$} & \rqhead{$\tau$} & \rqhead{$\Delta_2$} & \rqhead{$\Delta_{3:4}$} & \rqhead{$\tau$}\\
\midrule
Global shuffle & 0.012 & \textless{}0.001 & 1.008 & 0.058 & 0.001 & 1.032 & 0.037 & \textless{}0.001 & 1.020\\
Within-L1 shuffle & 0.477 & 0.006 & 1.374 & 0.411 & 0.010 & 1.239 & 0.421 & 0.010 & 1.249\\
L1 + category & 0.433 & 0.050 & 1.413 & 0.347 & 0.073 & 1.283 & 0.348 & 0.083 & 1.302\\
\rowcolor{adalignrow}\textbf{Semantic} & 0.399 & \textbf{0.084} & \textbf{1.445} & 0.320 & \textbf{0.100} & \textbf{1.305} & 0.321 & \textbf{0.110} & \textbf{1.328}\\
\textit{$\Delta$ vs.\ category} & \textcolor{rq2accent}{\(\downarrow\)\,\textbf{0.035}} & \textcolor{rq2accent}{\(\uparrow\)\,\textbf{0.035}} & \textcolor{rq2accent}{\(\uparrow\)\,\textbf{0.032}} & \textcolor{rq2accent}{\(\downarrow\)\,\textbf{0.027}} & \textcolor{rq2accent}{\(\uparrow\)\,\textbf{0.027}} & \textcolor{rq2accent}{\(\uparrow\)\,\textbf{0.022}} & \textcolor{rq2accent}{\(\downarrow\)\,\textbf{0.027}} & \textcolor{rq2accent}{\(\uparrow\)\,\textbf{0.027}} & \textcolor{rq2accent}{\(\uparrow\)\,\textbf{0.026}}\\
\midrule
\rowcolor{logicgray}
\textit{(c) Evidence} & \rqhead{Pairs} & \rqhead{TV} & \rqhead{Peak diff.} & \rqhead{Pairs} & \rqhead{TV} & \rqhead{Peak diff.} & \rqhead{Pairs} & \rqhead{TV} & \rqhead{Peak diff.}\\
\midrule
Binary & 2,134 & \textbf{0.344} & \cellcolor{adalignrow!70}\textbf{37.7\%} & 467 & \textbf{0.232} & \cellcolor{adalignrow!70}\textbf{16.7\%} & 579 & \textbf{0.253} & \cellcolor{adalignrow!70}\textbf{19.0\%}\\
Graded & 2,134 & \textbf{0.347} & \cellcolor{adalignrow!70}\textbf{38.1\%} & 467 & \textbf{0.234} & \cellcolor{adalignrow!70}\textbf{16.7\%} & 579 & \textbf{0.255} & \cellcolor{adalignrow!70}\textbf{19.7\%}\\
\midrule
\rowcolor{logicgray}
\textit{(d) Replication} & Within & Cross & Gap & Within & Cross & Gap & Within & Cross & Gap\\
\midrule
Binary & \cellcolor{adalignrow!70}\textbf{0.315} & 0.386 & \textcolor{rq2accent}{\textbf{+0.071}} & \cellcolor{adalignrow!70}\textbf{0.184} & 0.242 & \textcolor{rq2accent}{\textbf{+0.058}} & \cellcolor{adalignrow!70}\textbf{0.194} & 0.273 & \textcolor{rq2accent}{\textbf{+0.079}}\\
Graded & \cellcolor{adalignrow!70}\textbf{0.317} & 0.388 & \textcolor{rq2accent}{\textbf{+0.071}} & \cellcolor{adalignrow!70}\textbf{0.186} & 0.245 & \textcolor{rq2accent}{\textbf{+0.058}} & \cellcolor{adalignrow!70}\textbf{0.195} & 0.274 & \textcolor{rq2accent}{\textbf{+0.079}}\\
\bottomrule
\end{tabularx}
\endgroup
\caption{Relevance ambiguity across SID resolutions: (a) graded entropy and
ambiguous mass; (b) hierarchy controls; (c) heterogeneity under exact E/S/C
matching; and (d) replication across disjoint positive subsets.}
\label{tab:rq2-resolution}
\end{table}

Relevance ambiguity is concentrated at coarse SID depths, with substantial
variation across queries and identifier assignments.
\mbox{Table~\ref{tab:rq2-resolution}(a)} reports nonzero refinement entropy in
68.8\%--78.9\% of queries at L1, falling to 8.5\%--18.0\% at L4.
Semantic SIDs allocate more entropy to L3--L4 than the controls in
\mbox{Table~\ref{tab:rq2-resolution}(b)}. Their mean entropy-weighted depth
exceeds the L1 + full-category control by $0.022$--$0.032$
($p_{\mathrm H}=0.024$ after Holm correction). This comparison holds tree
structure, relevance judgments, L1 prefixes, and category paths fixed.
Queries with identical E/S/C counts and total SID entropy exhibit distinct
depth profiles. Under graded relevance, matched profiles in
\mbox{Table~\ref{tab:rq2-resolution}(c)} differ by
0.234--0.347 in mean total variation, and the depth of maximum ambiguity
differs in 16.7\%--38.1\% of pairs.
In \mbox{Table~\ref{tab:rq2-resolution}(d)}, graded profiles from disjoint
subsets of the same query have 18.2\%--28.8\% smaller distances than
profiles from matched queries ($p_{\mathrm H}<0.001$). Binary relevance yields
the same ordering. Definitions and sensitivity analyses appear in
Appendix~\ref{app:rq2-resolution}.

\subsubsection{Measures and controls}
\label{app:rq2-resolution}

\paragraph{Population and measures.}
The analysis uses the retrieval-eligible test sets: 6,014 US, 1,656 ES,
and 1,883 JP queries. Duplicate products are merged before normalization,
retaining the highest observed grade. Binary evidence gives every E/S/C
positive unit weight; graded evidence uses weights 3, 2, and 1. The subsets
with multiple positives contain 4,649, 1,348, and 1,534 queries, all with
nonzero observed SID entropy under the semantic hierarchy.

Refinement entropies are measured in nats. $Q$ is the percentage of queries
with nonzero refinement entropy; $M$ is the mean percentage of relevance
mass at parents with multiple positive children.
For nonzero total entropy,
$r_\ell(q)=\Delta_\ell(q)/\sum_j\Delta_j(q)$ gives its allocation
across depths and $\tau(q)=\sum_\ell\ell\,r_\ell(q)$ its mean location.
$\Delta_{3:4}=\Delta_3+\Delta_4$ summarizes fine-resolution ambiguity.

Arrows compare semantic assignments with the L1 + full-category control.
TV measures profile distance, and Peak diff. denotes pairs with disjoint
maximizing depths. Within and Cross compare disjoint positive subsets
from the same query and from matched queries, with Gap equal to Cross minus
Within under identical grade counts and total SID entropy.

The capacity-normalized statistic is
\begin{equation}
\bar\Delta_\ell(q)=
\frac{\Delta_\ell(q)}{
 \sum_{u\in\mathcal P_{\ell-1}}\mu_{\ell-1}(u\mid q)
 \log|\mathrm{Ch}(u)|},
\label{eq:rq2-normalization}
\end{equation}
with value zero when the denominator is zero. The denominator includes
every catalog-legal child of each relevance-bearing parent. Raw entropy measures unresolved relevance, while normalized entropy
expresses it relative to the available branching capacity. Prefix projection
preserves parent mass and satisfies the entropy chain rule.

The relevance mass in Table~\ref{tab:rq2-resolution}(a) is the query-mean
quantity $M_\ell(q)=\sum_u\mu_{\ell-1}(u\mid q)
\mathbf{1}[|\{v\in\mathrm{Ch}(u):\mu_\ell(v\mid q)>0\}|>1]$.
It measures relevance mass at parents with multiple positive children.
$Q$ counts queries with at least one branching prefix, and $M$ weights
branching prefixes by their relevance mass. Both statistics include all
retrieval-eligible queries.

\paragraph{Prevalence and magnitude.}
Table~\ref{tab:rq2-resolution}(a) reports ambiguity prevalence and raw
entropy, which decreases with depth in all three locales. The rise in
capacity-normalized entropy from L3 to L4 reflects reduced mean logarithmic
branching capacity: from 1.858 to 0.306 nats in US, 1.474 to 0.348 in ES,
and 1.464 to 0.331 in JP.

\paragraph{Topology-preserving randomization.}
Global shuffling reassigns the catalog SID multiset to items while
preserving tree topology and SID multiplicities. Within-L1 shuffling performs the same
reassignment separately within each original first-level prefix. The second
control preserves every query's L1 relevance distribution exactly while
disrupting finer semantic organization. Both controls preserve query positive
counts, item relevance frequencies, and evidence weights. Each control uses
999 assignments per locale. The primary statistic is mean entropy-weighted depth $\tau(q)$ among
queries with positive total SID entropy. SID collisions after reassignment
can produce zero total entropy. Relative to the semantic assignment,
within-L1 shuffling changes the eligible query count by at most one per
locale and assignment; global shuffling preserves this count.

Table~\ref{tab:rq2-controls} reports observed query-bootstrap confidence
intervals alongside the central 95\% of the random-assignment distribution.
The permutation envelopes describe variation across identifier assignments;
query-bootstrap intervals describe variation across queries. Two-sided permutation $p$-values use twice
the smaller empirical tail probability, with the plus-one correction.
Holm correction covers twelve contrasts across three locales and four
shuffle controls: global, within-L1, within-L1 with the first three category
levels fixed, and within-L1 with the complete category path fixed.
The reported depth contrasts have $p_{\mathrm H}=0.024$.
Relative to within-L1 shuffling, semantic SIDs reduce L2 entropy by
0.0783, 0.0906, and 0.1000 nats in US, ES, and JP, respectively, and
increase entropy at L3 and L4. Semantic organization places more of the
relevance branching at finer refinements.

\begin{table}[!htbp]
\centering\appendixtablestyle
\begin{tabular*}{\linewidth}{@{\extracolsep{\fill}}lrrr@{}}
\toprule
Hierarchy & ESCI-US & ESCI-ES & ESCI-JP\\
\midrule
Semantic & 1.4449\,[1.4327, 1.4569] & 1.3054\,[1.2850, 1.3259] & 1.3276\,[1.3097, 1.3477]\\
Global shuffle & 1.0082\,[1.0070, 1.0097] & 1.0325\,[1.0286, 1.0369] & 1.0204\,[1.0177, 1.0234]\\
Within-L1 shuffle & 1.3745\,[1.3735, 1.3756] & 1.2388\,[1.2375, 1.2405] & 1.2490\,[1.2478, 1.2507]\\
L1 + full category & 1.4128\,[1.4101, 1.4158] & 1.2831\,[1.2793, 1.2870] & 1.3021\,[1.2982, 1.3064]\\
\bottomrule
\end{tabular*}
\caption{Entropy-weighted depth $\tau$ under topology-preserving controls.
Semantic brackets give query-bootstrap 95\% confidence intervals; control
brackets give the central 95\% of 999 identifier assignments.
The full-category control preserves both L1 and the complete category path.
Means of $\tau$ are computed conditional on positive total SID entropy.}
\label{tab:rq2-controls}
\end{table}

\paragraph{Positive-count and evidence sensitivity.}
For each multi-positive query, we uniformly sample two distinct positives,
preserve their relevance weights, and average the depth profile over 30
subsamples. Under binary and graded evidence, mean raw entropy ranges from
$0.4835$ to $0.5916$ nats at L1 and from $0.0052$ to $0.0076$ at L4.
The concentration at early depths persists with positive count fixed.

On identical query sets, binary and graded relevance differ by at most
$0.0031$ in mean normalized ambiguity, with query-level global ambiguity
correlations of $0.9986$--$0.9991$.

Observed mean intervals use 2,000 query-bootstrap draws; matched-pair
intervals resample disjoint pairs. Randomization and subsampling are fixed
independently of their results.

\paragraph{Controlling catalog taxonomy.}
We further restrict item-to-SID shuffling to groups sharing an L1 prefix and
the complete catalog category path. This control preserves category
occupancy of SID slots, the query's category composition, its first-level
relevance distribution, and the catalog SID multiset. The fractions of
positive item occurrences eligible for reassignment are 88.6\%, 76.5\%,
and 78.2\% in US, ES, and JP, respectively.
Semantic entropy-weighted depth exceeds this control in every
locale (Table~\ref{tab:rq2-controls}), with differences of $0.0320$,
$0.0224$, and $0.0256$, respectively.

\paragraph{Replication across disjoint positive subsets.}
For each query and grade, we split
$2\lfloor N_{\mathrm{grade}}/2\rfloor$ products into two disjoint halves
of equal size. Eligible queries have at least two positives per half and
distinct positive SIDs. Disjoint query pairs match the half-grade counts.
Within-query and cross-query comparisons have identical E/S/C
composition and total SID entropy. The analysis includes 1,244 US pairs
(2,488 queries), 379 ES pairs (758 queries), and 479 JP pairs (958 queries).

For a pair $(q,q')$, the within-query statistic averages
$\mathrm{TV}(r_{q,A},r_{q,B})$ and
$\mathrm{TV}(r_{q',A},r_{q',B})$; the cross-query statistic averages
$\mathrm{TV}(r_{q,A},r_{q',B})$ and
$\mathrm{TV}(r_{q',A},r_{q,B})$. We average 100 annotation splits within
each disjoint query pair. Confidence intervals use 2,000 pair-bootstrap
draws, and two-sided tests use 10,000 sign randomizations.
Within-query distances are lower under binary and graded evidence in all locales
(Table~\ref{tab:rq2-resolution}(d); six-test Holm-adjusted $p<0.001$).
The graded reductions relative to cross-query distance are 18.2\%, 23.8\%,
and 28.8\%. The smaller within-query distances show consistent depth
structure across disjoint subsets of each query's observed judgments.

\paragraph{Exact evidence composition.}
Table~\ref{tab:rq2-resolution}(c) pairs queries with identical E, S, and C
counts. We exclude queries with positive-SID collisions (5 US, 12 ES, and
1 JP query) and form disjoint pairs within each count stratum. The analysis
contains 2,134 US, 467 ES, and 579 JP pairs. Equal grade counts and distinct
positive SIDs give equal total SID entropy under both evidence mappings.
At fixed relevance composition and total SID entropy, queries exhibit
distinct depth profiles under both binary and graded relevance.
Profile TV compares normalized depth profiles. Dominant depths differ when
the sets of entropy-maximizing depths are disjoint, with a tolerance of
$10^{-12}$ for tied maxima.

\paragraph{Restricting the relevance definition.}
On fixed subsets with at least two Exact positives (2,698 US, 858 ES, and
1,063 JP queries), mean refinement entropy remains nonzero at every depth
under E-only, E+S, and E+S+C evidence, with the original relevance weights
renormalized over the retained label set. With E alone, mean entropy ranges
from $0.7941$ to $1.2157$ nats at L1 and from $0.0226$ to $0.0381$ at L4,
showing that relevance branching persists under Exact-only evidence.
Appendix~\ref{app:training-robustness} reports retrieval results for
training with these relevance definitions.

\paragraph{Refinement entropy and supervision.}
Refinement entropy locates branching in the observed relevance distribution.
At a zero-entropy refinement, observed positive mass is concentrated on one
child. For a fixed local predictor, relevance-weighted one-hot sampling
and explicit conditional targets have the same expected gradient
(Eq.~\eqref{eq:gradient-variance}). Explicit targets remove the variance
due to target sampling. Training and scoring comparisons appear in
Appendices~\ref{app:training-controls}, \ref{app:rq3}, and~\ref{app:rq4}.

\subsection{Incomplete evidence and resolution limits}
\label{sec:rq6}

\begin{figure}[!htbp]
\centering
\includegraphics[width=0.85\linewidth,trim=0 0 0 296bp,clip]{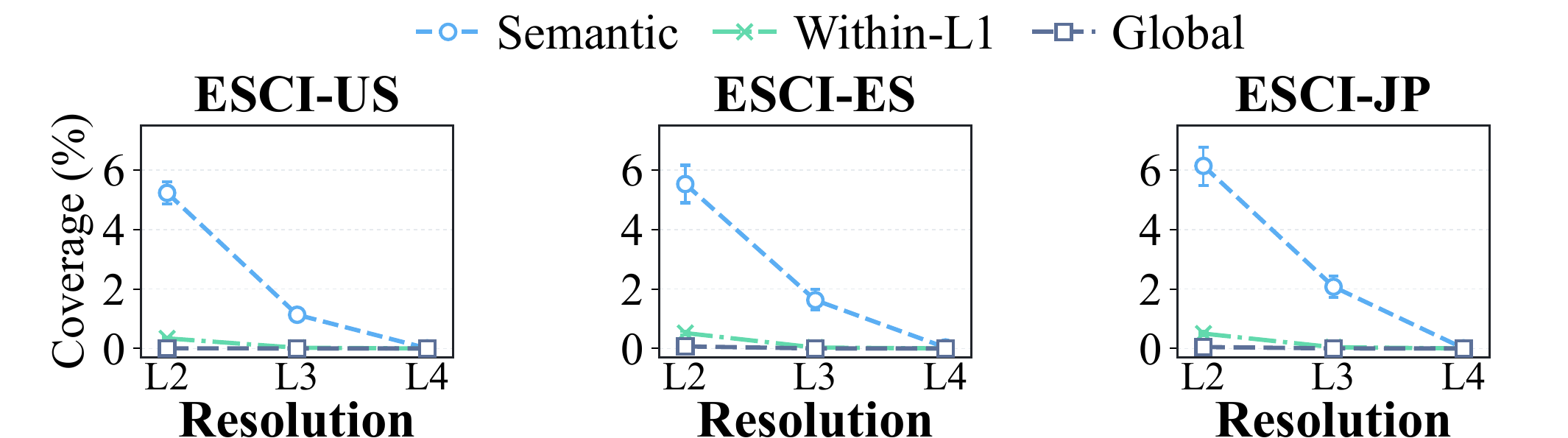}
\caption{Coverage of withheld relevance mass by SID depth at 25\% retention
under semantic identifiers, within-L1 reassignment, and global reassignment.
Coverage is evaluated over all products in supported prefixes.
Table~\ref{tab:rq6-sparse-evidence}(a) uses equal candidate counts across
identifier assignments. Error bars show 95\% confidence intervals.}
\label{fig:rq6-evidence-resolution}
\end{figure}

Using all observed positives as a reference, we examine how subsampling
changes target total variation (TV) and coverage of withheld relevance mass.
Figure~\ref{fig:rq6-evidence-resolution} compares semantic identifiers with
global and within-L1 reassignment at 25\% retention.
With candidate counts matched per query at 25\% retention,
Table~\ref{tab:rq6-sparse-evidence}(a)
reports expected L2 coverage of $2.56\%$--$3.12\%$ for semantic assignments
and $0.17\%$--$0.24\%$ after shuffling within L1.
The same ordering holds at L3,
with differences at both depths significant after Holm correction
($p_{\mathrm H}<0.004$).
Relative to uniform masking, preferentially retaining products with higher relevance
grades reduces mean TV, whereas concentrating evidence in fewer branches
increases it, as shown in Table~\ref{tab:rq6-sparse-evidence}(b).
Under uniform masking at 25\% retention, Table~\ref{tab:rq6-sparse-evidence}(c)
shows that $54.3\%$--$70.0\%$ of queries appear resolved
at L1 even though their full observed positive sets span multiple branches.
At L4, $70.7\%$--$72.9\%$ of relevance mass lacks retained support.
Incomplete judgments can produce apparent certainty at coarse depths
while missing relevant branches at finer depths. The subsampling design
and uncertainty estimates are detailed in Appendix~\ref{app:rq6-evidence}.
\begin{table}[!htbp]
\centering
\begingroup
\appendixtablestyle
\arrayrulecolor{black!65}
\setlength{\heavyrulewidth}{0.5pt}
\setlength{\lightrulewidth}{0.3pt}
\setlength{\tabcolsep}{3pt}
\renewcommand{\arraystretch}{1.04}
\noindent\begin{tabularx}{\linewidth}{@{}>{\raggedright\arraybackslash}X*{3}{>{\raggedleft\arraybackslash}p{0.14\linewidth}}@{}}
\toprule
\textbf{Evidence control} & \textbf{US} & \textbf{ES} & \textbf{JP}\\
\midrule
\rowcolor{adalignrow}[0pt][0pt]
\multicolumn{4}{@{}l@{}}{\strut\textbf{(a) Coverage at matched candidate counts (\%)}}\\
L2: Semantic & 2.64 & 2.56 & 3.12\\
L2: Within-L1 & 0.17 & 0.23 & 0.24\\
L3: Semantic & 0.46 & 0.69 & 0.98\\
L3: Within-L1 & 0.01 & 0.02 & 0.02\\
\addlinespace[1.0pt]
\rowcolor{adalignrow}[0pt][0pt]
\multicolumn{4}{@{}l@{}}{\strut\textbf{(b) Missingness: mean target TV}}\\
Uniform & 0.637 & 0.673 & 0.671\\
Grade-biased & 0.621 & 0.652 & 0.653\\
Branch-concentrated & 0.658 & 0.701 & 0.698\\
\addlinespace[1.0pt]
\rowcolor{adalignrow}[0pt][0pt]
\multicolumn{4}{@{}l@{}}{\strut\textbf{(c) Resolution limits (\%)}}\\
False resolution at L1 & 70.0 & 54.3 & 55.0\\
Unobserved L4 mass & 70.7 & 72.8 & 72.9\\
\bottomrule
\end{tabularx}\par
\arrayrulecolor{black}
\endgroup

\caption{Relevance recovery at 25\% positive retention with linear weights.
Coverage is measured at matched candidate counts; target TV is averaged
 over L1--L4.}
\label{tab:rq6-sparse-evidence}
\end{table}

\subsubsection{Evidence subsampling and statistical analysis}
\label{app:rq6-evidence}

\paragraph{Population and reference distribution.}
The evidence-recovery analysis uses the ESCI seen-query test partitions
from the CaLIR evaluation. It includes all queries with at least two
distinct positive products in the catalog. Eligibility depends on positive
count, independent of category routing, retrieval success, or ambiguity.
This population differs from the retrieval-eligible effectiveness sets.
Table~\ref{tab:rq6-population} accounts for all 10,177 test queries.
The population accounting lists single-positive queries separately;
retained--withheld comparisons use the multi-positive subset. Duplicate product judgments retain
the highest grade. Distinct products sharing a SID contribute their summed
mass after sampling. The benchmark's test grades define the reference
relevance distribution.

\begin{table}[!htbp]\centering\appendixtablestyle

\begin{tabular*}{\linewidth}{@{\extracolsep{\fill}}lrrrrrrrr@{}}\toprule
Locale & Source & Multi-pos. & Single-pos. & No positive & Catalog & Median $n_q$ & Retained & Actual (\%)\\\midrule
US & 6,137 & 4,742 & 1,395 & 0 & 288,372 & 4 & 1.47 & 29.3\\
ES & 1,877 & 1,418 & 321 & 138 & 101,957 & 7 & 2.37 & 27.2\\
JP & 2,163 & 1,594 & 367 & 202 & 119,052 & 7 & 2.29 & 27.1\\
\bottomrule\end{tabular*}
\caption{Population for the evidence-subsampling analysis. All queries with
at least two catalog-mapped positives are included, without category-routing
filters. No positive denotes queries without a catalog-mapped positive.
The final columns give the mean retained count and realized retention at
the nominal 25\% level.}
\label{tab:rq6-population}
\end{table}

The reference distribution normalizes relevance mass over all observed
positives for each query. Linear evidence assigns C/S/E weights $(1,2,3)$, binary evidence assigns weight one to
every positive, and exponential evidence assigns $(1,e,e^2)$. For each nominal
fraction $f$, the mask retains $\max(1,\lfloor f n_q\rfloor)$ products.
The 25\% budget consequently retains a mean of $1.47/2.37/2.29$ products and
realized mean fractions $29.3\%/27.2\%/27.1\%$ in US/ES/JP.
The realized fraction reflects integer rounding and the minimum of one
retained positive per query.

\paragraph{Experimental design.}
The exploratory design crosses three locales, five retention levels, and three identifier
assignments, with 20 paired replicates per condition. Each condition is
evaluated at four SID depths and at the individual-product level. Primary
comparisons use linear relevance weights, uniform masking, and 25\%
retention. Within each replicate, masking and identifier reassignment are randomized independently. Uniform masks sample nested positive sets without replacement.

Global and within-L1 controls permute entire SID paths over catalog products.
The multiset of paths is preserved, as are prefix sizes, numbers of legal
children, and SID collisions. Within-L1 assignment also preserves every
product's L1 prefix. The retained product set is identical across hierarchy
and evidence-weight comparisons. Partitioning by product identity is invariant to SID reassignment.

\begin{figure}[!htbp]
  \centering
  \includegraphics[width=\linewidth]{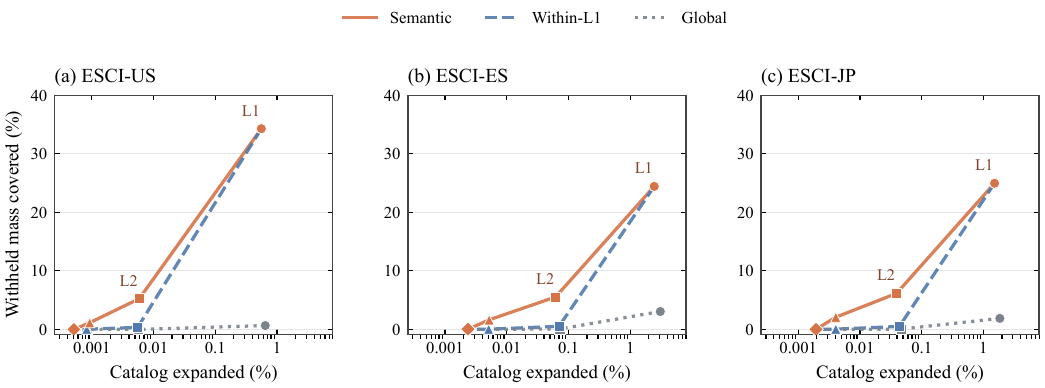}
  \caption{\textbf{Withheld coverage and catalog expansion at 25\% retention.}
  Curves connect L1--L4 under uniform masking and linear relevance weights.
  Circles, squares, triangles, and diamonds denote successive depths.
  The horizontal axis is logarithmic. Semantic and within-L1 assignments
  coincide at L1 by construction. Coverage counts withheld relevance mass
  sharing a prefix with retained positives.}
  \label{fig:rq6-cost}
\end{figure}

\paragraph{Measures and statistical inference.}
We measure target error by TV and Jensen--Shannon divergence in nats,
and quantify reference mass in prefix groups absent from the retained
evidence. Withheld coverage is the fraction of withheld positive mass in
groups containing at least one retained positive. Retained and withheld
products are disjoint. Withheld coverage is evaluated below 100\%
retention, where the withheld set is nonempty. Catalog expansion is the fraction of catalog
products contained in supported groups, measuring the size of the
resulting candidate set.
Figure~\ref{fig:rq6-cost} plots withheld relevance coverage against catalog
expansion across SID depths and identifier assignments.

Queries receive equal weight within each locale. The 20 paired replicates
are averaged within query before query-cluster bootstrap (4,000 draws)
and paired sign randomization (10,000 draws). The primary test family
compares semantic and global assignments on TV reduction and coverage gain
at all four SID depths in all three locales, using uniform masking, linear
weights, and 25\% retention. Holm correction is applied jointly to these
24 tests. Crossed-bootstrap intervals additionally resample queries and
replicates independently to account for variation in the finite set of
random assignments and masks. Table~\ref{tab:rq6-primary} reports the
complete set of primary comparisons. L4 coverage gains are
$0.0055/0.0349/0.0065$ points on US/ES/JP, with corresponding
Holm-adjusted $p$-values of $0.2476/0.0042/0.9969$.

\begin{table}[!htbp]\centering\appendixtablestyle
\begin{tabular*}{\linewidth}{@{\extracolsep{\fill}}llrrrr@{}}\toprule
Locale & Depth & TV reduction [95\% CI] & $p_{\rm H}$ & Coverage gain (pp) [95\% CI] & $p_{\rm H}$\\\midrule
US & L1 & 0.2320 [0.2256, 0.2382] & 0.0024 & 33.6331 [32.6809, 34.5729] & 0.0024\\
US & L2 & 0.0373 [0.0348, 0.0398] & 0.0024 & 5.2315 [4.8586, 5.6240] & 0.0024\\
US & L3 & 0.0083 [0.0073, 0.0095] & 0.0024 & 1.1337 [0.9796, 1.2986] & 0.0024\\
US & L4 & 0.0000 [0.0000, 0.0001] & 0.2476 & 0.0055 [0.0005, 0.0140] & 0.2476\\
ES & L1 & 0.1458 [0.1369, 0.1547] & 0.0024 & 21.3771 [20.0118, 22.7367] & 0.0024\\
ES & L2 & 0.0388 [0.0346, 0.0433] & 0.0024 & 5.4617 [4.8383, 6.1346] & 0.0024\\
ES & L3 & 0.0116 [0.0095, 0.0140] & 0.0024 & 1.6248 [1.2946, 1.9876] & 0.0024\\
ES & L4 & 0.0003 [0.0001, 0.0005] & 0.0042 & 0.0349 [0.0146, 0.0609] & 0.0042\\
JP & L1 & 0.1601 [0.1507, 0.1694] & 0.0024 & 23.0821 [21.7149, 24.4512] & 0.0024\\
JP & L2 & 0.0433 [0.0391, 0.0476] & 0.0024 & 6.0914 [5.4547, 6.7778] & 0.0024\\
JP & L3 & 0.0149 [0.0127, 0.0173] & 0.0024 & 2.0735 [1.7408, 2.4520] & 0.0024\\
JP & L4 & 0.0001 [0.0000, 0.0002] & 0.9969 & 0.0065 [0.0000, 0.0195] & 0.9969\\
\bottomrule\end{tabular*}
\caption{Primary comparisons of semantic and global identifier assignments
under linear relevance weights, uniform masking, and 25\% retention.
Positive differences favor semantic assignments. Brackets give pointwise crossed
query-and-replicate bootstrap 95\% intervals (4,000 draws). Adjusted
$p_{\rm H}$ values use 10,000 paired query sign randomizations with Holm
correction over all 24 tests.}
\label{tab:rq6-primary}
\end{table}

\begin{figure}[!htbp]
  \centering
  \includegraphics[width=\linewidth]{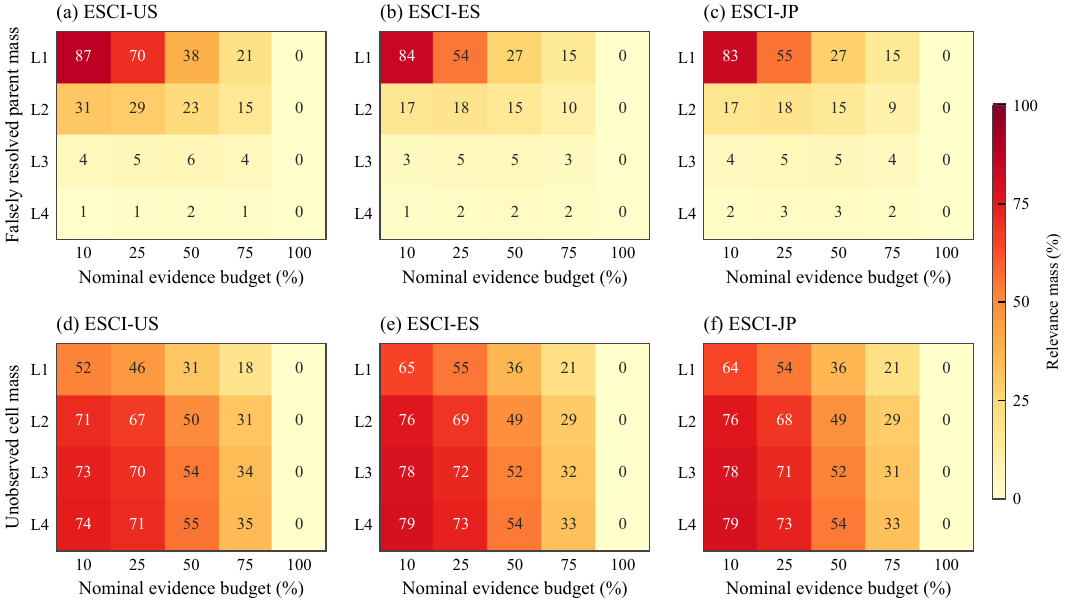}
  \caption{\textbf{Apparent resolution and missing support.} Results use
  semantic identifiers, uniform masking, and linear relevance weights.
  Top: reference mass on ambiguous parents that appear deterministic after
  subsampling. Bottom: reference mass in prefix groups with no retained
  positives. Percentages are rounded to integers.}
  \label{fig:rq6-certainty}
\end{figure}

\paragraph{False certainty and absent support.}
A parent is falsely resolved when its full observed positive set occupies
at least two children but its retained positives occupy exactly one. We
sum the reference mass of such parents. Missing support is the reference
mass in prefix groups with no retained positives
(Figure~\ref{fig:rq6-certainty}). These two quantities measure collapsed
branching and unobserved groups, respectively. At 25\% retention, $71\%$--$73\%$ of reference L4 mass belongs to
groups with no retained positives. Table~\ref{tab:rq6-strata} reports
results by positive-count stratum.

\begin{table}[!htbp]\centering\appendixtablestyle
\begin{tabular*}{\linewidth}{@{\extracolsep{\fill}}llrrrrrrr@{}}\toprule
Locale & $n_q$ & Queries & TV L1 & TV L2 & TV L3 & TV L4 & Cov. L1 & Cov. L2\\\midrule
US & 2-3 & 1,810 & 0.417 & 0.551 & 0.570 & 0.573 & 27.16 & 3.81\\
US & 4-7 & 1,756 & 0.571 & 0.770 & 0.794 & 0.801 & 28.82 & 3.89\\
US & 8-15 & 917 & 0.431 & 0.717 & 0.763 & 0.777 & 48.97 & 7.94\\
US & 16+ & 259 & 0.314 & 0.659 & 0.740 & 0.765 & 69.11 & 14.74\\
ES & 2-3 & 371 & 0.496 & 0.557 & 0.564 & 0.572 & 12.89 & 2.62\\
ES & 4-7 & 393 & 0.691 & 0.777 & 0.796 & 0.801 & 13.78 & 3.06\\
ES & 8-15 & 349 & 0.574 & 0.729 & 0.763 & 0.778 & 28.80 & 6.69\\
ES & 16+ & 305 & 0.450 & 0.688 & 0.745 & 0.764 & 47.18 & 10.93\\
JP & 2-3 & 423 & 0.494 & 0.552 & 0.564 & 0.574 & 13.66 & 3.73\\
JP & 4-7 & 452 & 0.667 & 0.778 & 0.795 & 0.805 & 17.32 & 3.46\\
JP & 8-15 & 394 & 0.562 & 0.727 & 0.765 & 0.778 & 30.30 & 6.87\\
JP & 16+ & 325 & 0.469 & 0.679 & 0.733 & 0.764 & 43.83 & 12.12\\
\bottomrule\end{tabular*}
\caption{Evidence recovery by positive-count stratum under the semantic
hierarchy, uniform masking, linear relevance weights, and 25\% retention.
Coverage is the percentage of withheld relevance mass recovered. Results
are shown for all four prespecified strata.}
\label{tab:rq6-strata}
\end{table}

\paragraph{Properties of relevance projection.}
Prefix aggregation preserves relevance mass and contracts TV and JS.
At 100\% retention, the estimated and reference distributions coincide.
SID permutations preserve catalog paths, within-L1 permutations preserve
L1 relevance distributions, and product-level quantities are invariant to
SID reassignment.

\paragraph{Invariance under restricted reassignment.}
A within-L1 permutation preserves $\pi_1(d)$ for every product. Both the
reference and masked L1 distributions, their support, and the expanded
catalog are identical to the semantic assignment. L1 serves as a control
for the preserved resolution.
At a one-to-one document partition,
the TV error is $1-\sum_{d\in S_q}\mu(d\mid q)$, where $S_q$ is the
retained positive set. This quantity is invariant to renaming unique
document identifiers. SID collisions account for small departures at L4.
Table~\ref{tab:rq6-full-reference} exhibits the corresponding equality at
L1 and near-equality at L4.

\begin{table}[!htbp]
\centering\appendixtablestyle

\begin{tabularx}{\linewidth}{@{}>{\raggedright\arraybackslash}Xlrrrrrr@{}}
\toprule
& & \multicolumn{4}{c}{Target TV $\downarrow$} & \multicolumn{2}{c}{L1 support (\%)}\\
\cmidrule(lr){3-6}\cmidrule(lr){7-8}
Locale & Assignment & L1 & L2 & L3 & L4 & Coverage & Expansion\\
\midrule
\rowcolor{adalignrow}US & Semantic & 0.4709 & 0.6702 & 0.6992 & 0.7075 & 34.28 & 0.56 \\
 & Within-L1 & 0.4709 & 0.7050 & 0.7073 & 0.7075 & 34.28 & 0.56 \\
 & Global & 0.7028 & 0.7075 & 0.7075 & 0.7075 & 0.65 & 0.65 \\
\midrule
\rowcolor{adalignrow}ES & Semantic & 0.5595 & 0.6884 & 0.7161 & 0.7275 & 24.43 & 2.45 \\
 & Within-L1 & 0.5595 & 0.7238 & 0.7275 & 0.7278 & 24.43 & 2.45 \\
 & Global & 0.7053 & 0.7272 & 0.7278 & 0.7278 & 3.05 & 3.07 \\
\midrule
\rowcolor{adalignrow}JP & Semantic & 0.5547 & 0.6851 & 0.7138 & 0.7287 & 24.96 & 1.51 \\
 & Within-L1 & 0.5547 & 0.7250 & 0.7285 & 0.7288 & 24.96 & 1.51 \\
 & Global & 0.7148 & 0.7284 & 0.7288 & 0.7288 & 1.88 & 1.85 \\
\bottomrule
\end{tabularx}
\caption{Target error and L1 support at 25\% retention. Semantic and
within-L1 assignments have identical L1 error, coverage, and expansion by
construction. Table~\ref{tab:rq6-sparse-evidence}(a) compares their affected
fine resolutions. L1 coverage and catalog expansion are percentages.}
\label{tab:rq6-full-reference}
\end{table}

\paragraph{Equal catalog budgets.}
To equalize candidate counts, let $C_h(q,r)$ be the supported catalog size
under assignment $h$ in replicate $r$, and set
$B(q,r)=\min_h C_h(q,r)$ across semantic, within-L1, and global assignments.
Uniformly sampling $B(q,r)$ products from each supported set gives expected
withheld coverage
$\widetilde{R}_h(q,r)=R_h(q,r)B(q,r)/C_h(q,r)$.
The common count depends only on catalog geometry and retained evidence;
withheld labels enter the coverage evaluation. The expectation is taken
over uniform samples from each supported candidate set.

At L2, the shared cap averages 8.12, 19.56, and 15.33 candidates per query
in US, ES, and JP. Semantic expected coverage is 2.64\%, 2.56\%, and
3.12\%, compared with 0.17\%, 0.23\%, and 0.24\% for within-L1
reassignment. Semantic identifiers retain higher fine-level coverage at
identical candidate counts within each query and replicate.

\setlength{\parskip}{0.5pc}
\clearpage

\section{Real-Product Case Studies}
\label{app:real-cases}

\begin{figure}[!htbp]
\centering
\includegraphics[width=\linewidth]{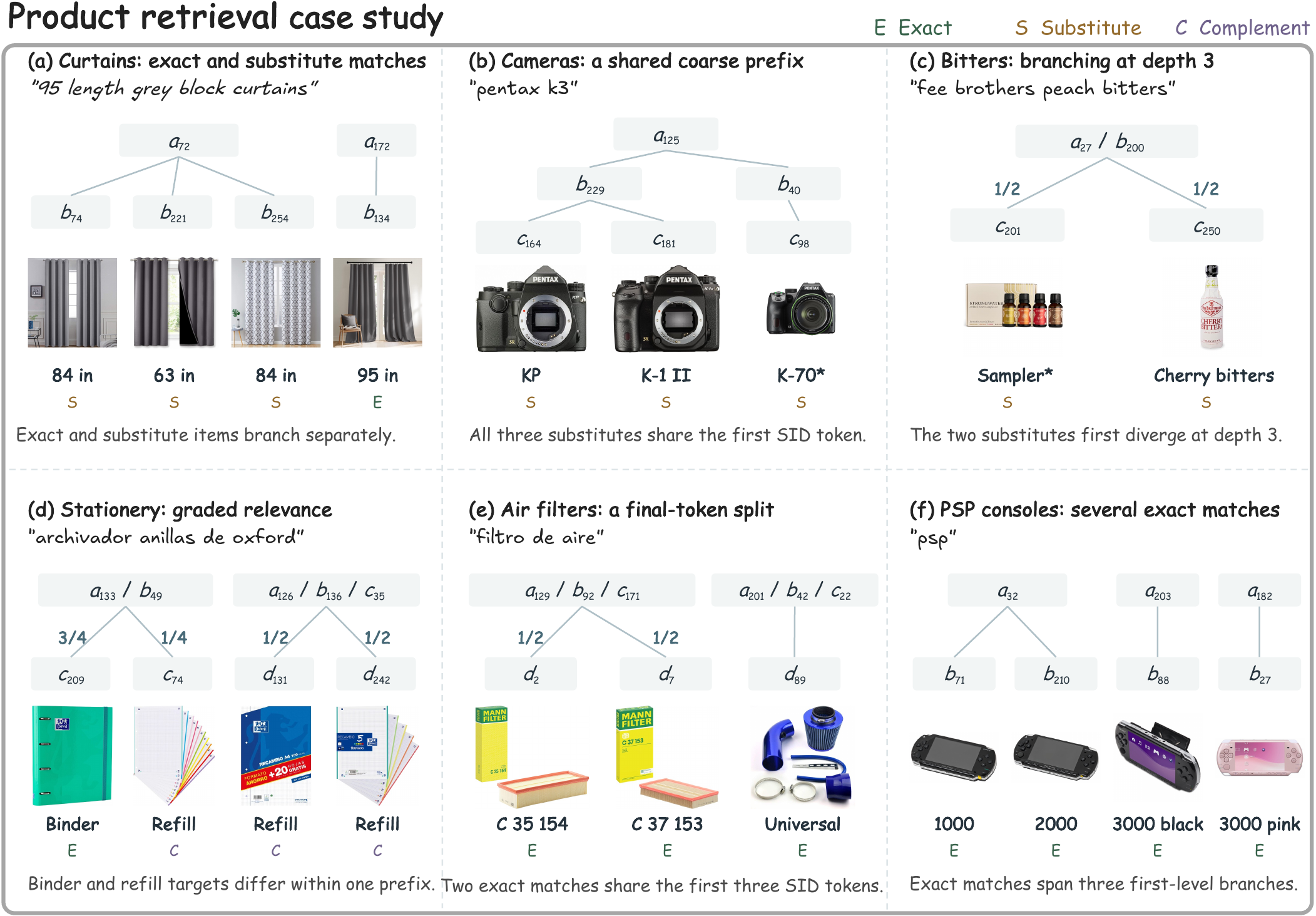}
\caption{\textbf{Relevance structure in product retrieval.}
Six ESCI queries exhibit different branching depths and local relevance
distributions in the SID hierarchy.
E, S, and C denote Exact, Substitute, and Complement judgments.
Fractions indicate conditional relevance mass over the displayed children.
Unary paths are compressed, and identifier-uniqueness suffixes are omitted.
For starred products, the K-70 image includes a lens although the catalog
listing is body-only, and the sampler image shows a four-bottle pack for a
historical five-bottle listing.}
\label{fig:real-cases}
\end{figure}

Figure~\ref{fig:real-cases} presents six ESCI queries with their judged
products and SID prefixes.
The examples highlight two aspects of multiresolution relevance: the depth at
which relevant products separate in the SID hierarchy, and the distribution of
relevance mass among competing children.
Across queries, relevant products may diverge immediately at coarse levels or
remain grouped until the final refinement, producing distinct supervision
patterns along the same type of identifier hierarchy.

\textbf{Branching depth.}
The examples show branching structures.
For the curtain query in (a), the shorter Substitute products fall under
$a_{72}$, while the Exact 95-inch product belongs to $a_{172}$, so relevance
separates at the first SID level.
The four Exact PSP matches in (f) likewise span three first-level branches,
showing that products with the same relevance grade need not share a coarse
prefix.
Other queries retain shared structure for several refinements.
The camera substitutes in (b) share $a_{125}$ before separating, while the two
bitters products in (c) share the first two SID tokens and diverge at depth 3.
The C~35~154 and C~37~153 air filters in (e) share three tokens and separate
only at the final SID token, while the Universal filter occupies another
coarse branch.
Thus, the resolution at which relevance branches is query-dependent, ranging
from the first decision to the final refinement.

\textbf{Conditional relevance mass.}
Branching depth alone does not determine the supervision target.
The displayed targets use E/S/C weights of $3/2/1$, normalized over positive
children at each parent.
In (d), the Exact binder and Complement refill below
$a_{133}/b_{49}$ receive conditional masses $3/4$ and $1/4$, reflecting their
different relevance grades within the same local decision.
The two Complement refills below $a_{126}/b_{136}/c_{35}$ each receive mass
$1/2$.
Equal local targets also arise for the two Substitute products in (c) and the
two Exact filters in (e).
SID structure determines which relevant products compete at a refinement,
while relevance grades determine their relative mass.
The same grade may occur in different branches, as in the PSP and curtain
examples, while different grades may share a parent, as in the stationery
example.
Thus, document-level relevance induces different local target distributions
across SID resolutions.
Table~\ref{tab:main-results} evaluates the effect of learning this structure
at scale.

\clearpage

\end{document}